\documentclass[PRD,aps,twocolumn,showpacs,preprintnumbers,amsmath,amssymb,shperscriptaddress,nofootinbib]{revtex4-1}
\usepackage[dvips]{graphicx}
\usepackage{epsf}
\usepackage{amsmath}
\usepackage{amssymb}
\usepackage{amsfonts}
\usepackage{times}
\usepackage[usenames]{color}
\usepackage[normalem]{ulem}
\usepackage[T1]{fontenc}
\usepackage[dvipsnames]{xcolor}
\usepackage{floatrow}

\usepackage{graphicx}
\usepackage{dcolumn}
\usepackage{bm}
\usepackage{color}

\begin{document}

\title{Simulating the scattering of low-frequency Gravitational Waves by compact objects using the 
finite element method}

\author{Jian-hua He}
\email[Email address: ]{hejianhua@nju.edu.cn}
\affiliation{School of Astronomy and Space Science, Nanjing University, Nanjing 210093, P. R. China}
\begin{abstract}
We investigate the wave effects of gravitational waves (GWs) using numerical simulations with the finite element method (FEM) based on the publicly available code {\it deal.ii}. We robustly test our code using a point source monochromatic spherical wave. We examine not only the waveform observed by a local observer but also the global energy conservation of the waves. We find that our numerical results agree very well with the analytical predictions. Based on our code, we study the scattering of GWs by compact objects. Using monochromatic waves as the input source, we find that if the wavelength of GWs is much larger than the Schwarzschild radius of the compact object, the amplitude of the total scattered GWs does not change appreciably due to the strong diffraction effect, for an observer far away from the scatterer. This finding is consistent with the results reported in the literature. However, we also find that, near the scatterer, not only the amplitude of the scattered waves is very large, comparable to that of the incident waves, but also the phase of the GWs changes significantly due to the interference between the scattered and incident waves. As the evolution of the phase of GWs plays a crucial role in the matched filtering technique in extracting GW signals from the noisy background, our findings suggest that wave effects should be taken into account in the data analysis in the future low-frequency GW experiments, if GWs are scattered by nearby compact objects in our local environment. 
\end{abstract} 

\maketitle
\section{Introduction}
The landmark discovery of gravitational waves~\cite{Abbott:2016blz}, ripples in the fabric of spacetime trigged by the merger of two black holes, confirms a major prediction of Einstein's general relativity. It marks the climax of century-long speculation and decades-long painstaking work for hunting such waves. The direct detection of gravitational waves ushered us into a new era of astronomy, allowing us to probe the Universe in an unprecedented way. 

In the coming decades, ground- and space-based GW experiments will continue to study GWs, such as the Einstein Telescope (ET)~\cite{Einstein_tele}, 40-km LIGO~\cite{LIGO40}, eLISA~\cite{eLISA}, DECIGO~\cite{Sato_2009}, and Pulsar Timing Arrays (PTA)~\cite{2010CQGra..27h4013H}. The range of the frequency of GWs will be explored from ultra-low frequency regimes such as $10^{-9} {\rm Hz}-10^{-8} {\rm Hz}$ (PTA), $10^{-4}{\rm Hz}-10^{-1} {\rm Hz}$ (eLISA) ,and $10^{-2}{\rm Hz}-1 {\rm Hz}$ (DECIGO) to high frequency regimes such as $10{\rm Hz}-10^{4} {\rm Hz}$ (ET and 40-km LIGO). These experiments will usher us into an era of routine GW astronomy and enable us to study the GW phenomenon in unprecedented detail. The combination of these experiments promises to address a variety of outstanding problems in astrophysics, cosmology, and particle physics. 

The low-frequency GWs observed by space-borne GW interferometer may shed light on the population of supermassive BHs (SMBHs) with a mass $>10^6M_{\odot}$~\cite{Baibhav:2019rsa}, which is believed to ubiquity reside in the centers of galaxies~\cite{Bellovary:2015ifg,Bartos:2016dgn,Stone:2016wzz}. SMBHs are major cosmic players and are important to galaxy formation~\cite{Haehnelt:2000sx}. The associated jets and accretion disks are the most energetic phenomena in the Universe, which often outshine the entire host galaxy. Therefore, their properties play an important role in the dynamics of stars and gas at galactic scales~\cite{Ferrarese:2000se}. However, the details of the birth and growth of SMBHs are not well understood yet. The observation of low-frequency GWs, such as in the millihertz band, can probe the spacetime geometry of black hole binaries and their surroundings, which opens a window that can inform us of the properties of these BHs throughout cosmic history. GW observations, thus, can help us to answer many fundamental questions, as to, where and when do the first massive black holes form, how do they grow and assemble, and how their properties are connected to the properties of their host galaxies?

However, in order to achieve these scientific goals, it is crucial to accurately infer the physical properties of the distant source from the observed GW signals. This process, however, is likely beset by the wave nature of GWs. Unlike the visible light, the wavelength of the low-frequency GWs is much larger. In the millihertz band, for example, the wavelength can be as large as several times the astronomical unit (${\rm AU}$). When the wavelength of the GWs is comparable to the Schwarzschild radius of an object, its propagation no longer obeys the geometrical optics. Further, if the gravitational waves come from the same compact binaries, they should be coherent and interference. In all these cases, the wave effects of GWs are significant and have to be taken into account~\cite{Bontz,Ohanian:1974ys}. 

The wave effects in a lensing system have already been studied in the previous work~\cite{PhysRevD.34.1708,Meena:2019ate,Deguchi,Schneider,Ruffa_1999,DePaolis:2002tw,Takahashi:2003ix,1999PThPS.133..137N,Suyama:2005mx,Christian:2018vsi,DOrazio:2019ens,Zakharov_2002}. In these pioneer work (e.g. Ref.~\cite{Schneider}), the lensing system is usually considered as a thin lens model: the gravitational field of the compact lens is weak, and the deflection angle due to the lens masses is small. The impact parameter of the incident rays is thought to be much larger than the Schwarzschild radius of the lens mass. Thus, the changes in the amplitude of the incident waves are negligible at the location of the lens mass. The changes in the amplitude, however, are mainly due to the deflection near the lens, which is small at the beginning but becomes significant after the long journey of the light rays from the lens to the observer (see discussions in (4.86) in Ref.~\cite{Schneider}). 

In addition to the thin-lens model, the previous work also applied {\it Kirchoff's} diffraction theory to the gravitational lensing system~\cite{Schneider}. The lens is considered as an optic aperture. The lensed signal is then thought to be the diffraction of the waves from the aperture. However, it should be noted that an important assumption in {\it Kirchoff's} theory is that boundaries outside the aperture are set to be zero. {\it Kirchoff's} theory also assumes that only waves re-radiated from the aperture can be observed by the observer. The diffraction integral in {\it Kirchoff's} theory, indeed, only considers the scattered waves, which neglects the contribution of the original incident waves~\cite{Kirchoff} (see Eq.(5) in Ref.~\cite{Takahashi:2005sxa}). This is valid in the system of wave optics, as in this case most of the original incident waves are blocked by obstacles outside the aperture. However, if the amplitude of the scattered wave is comparable to that of the incident wave, the original incident waves have to be considered as well. 

\begin{figure}
{\includegraphics[width=\linewidth]{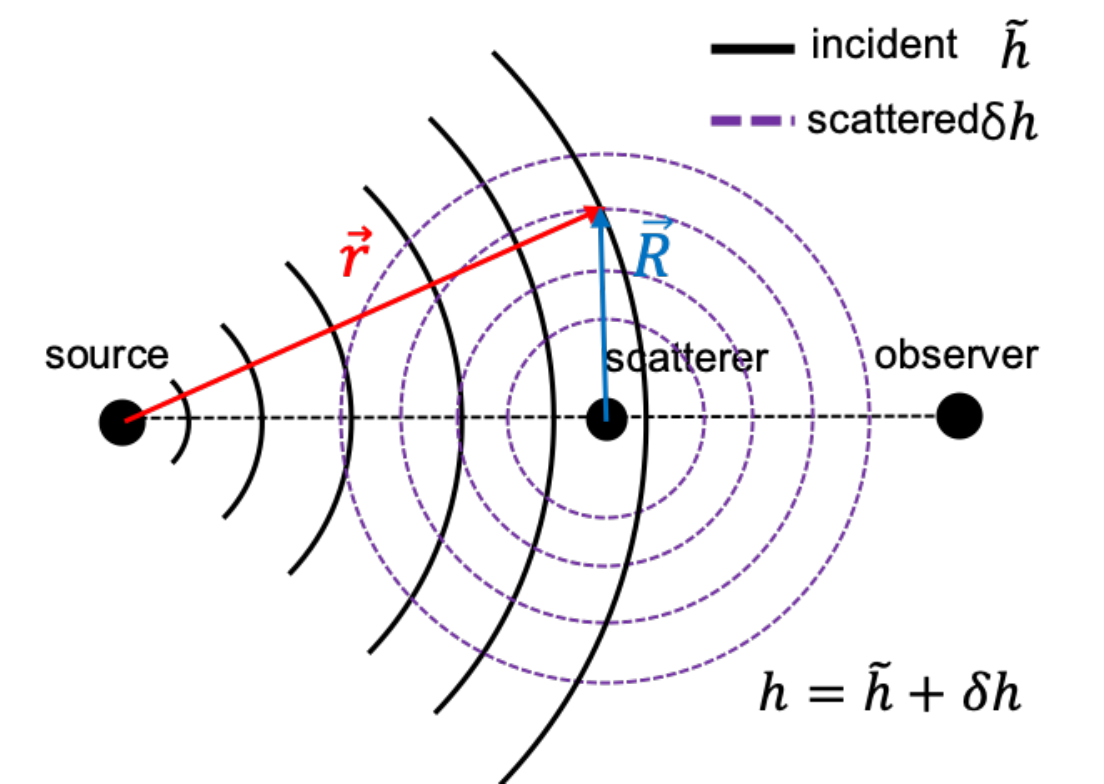}}
\caption{The schematic of our scattering model. The model is assumed to be in an infinte space. The GW signal is scattered by a compact object (scatterer) in the way between the source and the observer. The total scattered wave $h$ is the interference of the incident wave $\tilde{h}$ and the scattered wave $\delta h$, namely, $h=\tilde{h}+\delta h$. \label{figone}}
\end{figure} 

A more mathematically consistent way to describe the lensing system is to treat it as a scattering system~\cite{GR_steven_weinberg,Peters,Takahashi:2005sxa,Sorge:2015yoa}. The total scattered wave $h$ is the interference of the incident wave $\tilde{h}$ and the scattered wave $\delta h$ (as illustrated in Fig.~\ref{figone}). An advantage of this scheme is that it can naturally reproduce the no-scatterer limit, namely, if there were no scatterer, there would be only the original incident waves ({\it Kirchoff's} diffraction integral fails to give this point, even with an infinite large aperture). Indeed, the scattering of waves by potential wells has been well studied in Quantum Mechanics. As shown in Ref.~\cite{Peters}, the wave equation of GWs propagating in a non-uniform background spacetime is similar to that of the well-known Schr{\"o}dinger's equation for the scattering of two charged particles. Moreover, Refs.~\cite{Peters,Takahashi:2005sxa} also generalized the scattering model that uses a plane wave as the incident wave to a spherical one. They obtained a formal expression for the scattered waves in terms of Green's function. 

This work aims to extend the previous work to more general cases using numerical simulations. Unlike the previous work that focuses on the frequency-domain, our analyses are in the time-domain. The advantage of the time-domain is that it is directly related to the waveform of GWs. Accurately predicting the waveform, in particular the phase of GWs, plays a crucial role in using the matched filtering technique (see, e.g. ~\cite{Allen:2004gu}) to dig out GW signals from the very noisy background data (See e.g.~\cite{Maggiore:1999vm} for a review).

For the numerical implement, we adopt the finite element method (FEM). The FEM is an effective and well-developed method for numerically solving partial differential equations (see e.g. text book~\cite{FEMbook} for details). Unlike most of the conventional methods, such as the finite difference method (FDM), the FEM is based on the weak formulation of PDEs. The basic idea of the FEM is to present the domain of interest as an assembly of finite elements. On each finite element, the solution of PDEs is approximated by local shape functions. A continuous PDE problem thus can be transformed into a discretized finite element with unknown nodal values. These unknowns form a system of linear algebraic equations, which can then be solved numerically. Compared to the conventional numerical methods, such as the FDM, which requires regular meshes, the advantage of the FEM is that it can work for complex boundaries. It can also provide very good precision even with simple approximation shape functions. Further, due to the locality of the approximation shape functions, the resulting linear algebraic equations are sparse, which can be solved efficiently in the numerical process.

Throughout this paper, we adopt the geometric unit $c=G=1$, in which $1\,{\rm Mpc}=1.02938\times 10^{14} {\rm Hz}^{-1}$ and $1 M_{\odot}=4.92535\times 10^{-6} {\rm Hz}^{-1}\,.$ The advantage of this unit system is that time and space have the same unit. 

This paper is organized as follows: In Section~\ref{WF}, we introduce the Strong and Weak formulation of the wave equation. In Section~\ref{FEM}, we describe the FEM for wave equations. In Section~\ref{CT}, we present several tests of our code using spherical waves. In Section~\ref{GWP}, we simulate GWs scattered by compact objects. In Section~\ref{cl}, we summarize and conclude this work.

\section{Strong and Weak formulation of the wave equation\label{WF}}
In this section, we present an overview of the mathematical basis of the wave equation. To begin with, we discuss the {\it strong formulation} of the wave equation and then we will introduce the {\it weak formulation}.
\subsection{strong formulation}
Let $\Omega \subset \mathbb{R}^3$ be a bounded domain with boundary $\partial \Omega$. The boundary value problem (BVP) we aim to study in this work can be presented as
\begin{equation}
c^2\nabla^2 u -\frac{\partial^2}{\partial t^2}u =c^2f\quad {\rm in} \quad \Omega\times (0,T] \label{strong}\,,
\end{equation}
which is a second-order linear hyperbolic partial differential equation (PDE) 
with initial conditions
\begin{align}
u(x,0) = u_0(x) \quad {\rm in} \quad \Omega  \,. \label{S1}
\end{align}
Let $\partial \Omega_1$ and $\partial \Omega_2$ are subsets of the boundary $\partial \Omega$ with $\partial \Omega_1\cup \partial \Omega_2=\partial \Omega$ and $\partial\Omega_1\cap\partial\Omega_2=\emptyset$. 
We assume that the boundary conditions on $\partial \Omega_1$ are Dirichlet and on $\partial \Omega_2$ is Neumann
\begin{align}
u(x,t) &= u_1(x,t) &{\rm on}& \quad \partial\Omega_1\times (0,T]  \label{S2}\\ 
\hat{n}\cdot\nabla u & = q(x,t) &{\rm on}& \quad \partial\Omega_2\times (0,T] \label{S3} \,, 
\end{align}
where $\hat{n}$ is the outward pointing unit vector normal to $\partial \Omega_2$. $q(x,t)$ is a given function on the boundary $\partial\Omega_2$.

In mathematics, Eq.~(\ref{strong}), together with the initial and boundary conditions Eqs.~(\ref{S1},\ref{S2},\ref{S3}), are called the {\it strong formulation}. The {\it strong formulation} is the most commonly used format of the wave equation. It can be numerically solved using some straightforward methods such as the finite difference method (FDM). However, the {\it strong formulation} is not the only format of the wave equation. In what follows, we introduce an alternative format, namely, the {\it weak formulation} of the above equations. 
\subsection{Weak formulation}
Multiply Eq.~(\ref{strong}) by {\it a test function} $\phi$ and then integrate over $\Omega$, we obtain a variational equation
\begin{widetext}
\begin{align}
&-\int_{\Omega}\nabla u \cdot \nabla (c^2\phi)\,\mathrm{d}x + \int_{\partial \Omega}(\hat{n}\cdot\nabla u) (c^2\phi)\,\mathrm{d}x -\frac{\partial^2}{\partial t^2}\int_{\Omega} u \phi \, \mathrm{d}x =\int_{\Omega} c^2f\phi \,\mathrm{d}x  \quad \forall \phi \in V\,,\label{weak}
\end{align} 
\end{widetext}
where we have used the Green's formula
\begin{equation}
\int_{\Omega} c^2\phi \nabla^2 u \,\mathrm{d}x= -\int_{\Omega}\nabla(c^2\phi)\cdot\nabla u \,\mathrm{d}x + \int_{\partial \Omega}c^2\phi \hat{n}\cdot\nabla u \,\mathrm{d}x \,.\nonumber
\end{equation}
The second term on the right hand represents the boundary term, which can be further split according to different boundary conditions
\begin{align}
\int_{\partial \Omega}(\hat{n}\cdot\nabla u) (c^2\phi)\,\mathrm{d}x &=\int_{\partial \Omega_1}(\hat{n}\cdot\nabla u) (c^2\phi)\,\mathrm{d}x \nonumber \\
&{}+ \int_{\partial \Omega_2}(\hat{n}\cdot\nabla u) (c^2\phi)\,\mathrm{d}x \,.
\end{align}

The test function $\phi$ is chosen in such a way that it vanishes on the subset of boundaries with Dirichlet boundary conditions ${\partial \Omega_1}$, which is equivalent to say that we restrict $\phi$ to lie in 
\begin{equation}
V:=\{\phi:\phi\in H^{1}(\Omega),\phi|_{\partial \Omega_1}=0\}\,,\label{Vspace}
\end{equation}
where $H^{1}(\Omega)=W^{1,2}(\Omega)$ is called the first order {\it Sobolev space}, which is also a Hilbert space, meaning that $\phi$ and its first order weak derivatives $\partial_x \phi$ are square integrable. 
\begin{equation}
\left\|\phi\right\|_{H^1(\Omega)}=\left[\int_{\Omega}\sum_{|\alpha|\leq 1} |\partial^{\alpha}_x\phi(x)|^2dx\right]^{\frac{1}{2}}<\infty\,.\nonumber
\end{equation}

Given the restrictions of the test function $\phi$, the boundary term in Eq.~(\ref{weak}) can be reduced to
\begin{align}
\int_{\partial \Omega}(\hat{n}\cdot\nabla u) (c^2\phi)\,\mathrm{d}x &=\int_{\partial \Omega_2}(\hat{n}\cdot\nabla u) (c^2\phi)\nonumber \\
&=-\int_{\partial \Omega_2}\frac{\partial{u}}{\partial{t}} (c\phi) \,\mathrm{d}x \nonumber \quad\,.
\end{align}
Here, in the second equality, we have imposed an absorbing boundary condition
\begin{equation}
\hat{n}\cdot\nabla u=-\frac{1}{c}\frac{\partial{u}}{\partial{t}}\quad {\rm on} \quad \partial\Omega_2\times (0,T] \,.\label{boundary2}
\end{equation}
The physical meaning of the absorbing boundary condition will be explained later on.

The above we have discussed the boundary condition on $\partial \Omega_2$. On the other hand, the boundary condition on $\partial \Omega_1$,
\begin{equation}
\left \{
\begin{aligned}
u(x,t) &= u_0(x,t)   \\
\frac{\partial u(x,t)}{\partial t} &= \frac{\partial u_0(x,t)}{\partial t}  
\end{aligned}
\right. \quad {\rm on} \quad \partial\Omega_1\times (0,T] \,, \label{boundary1}
\end{equation}
is called the {\it essential boundary condition}, which does not affect the variational
equation Eq.~(\ref{weak}) but this condition must be imposed directly on the solution itself.

Equipped with the boundary conditions Eqs.~(\ref{boundary2},\ref{boundary1}), Eq.~(\ref{weak}) is called the {\it weak formulation} of Eq.~(\ref{strong}). According to the Lax-Milgram theorem~\cite{LaxMilg}, if Eq.~(\ref{weak}) holds for all $\phi\in V$, it has a unique solution $u\in V$. This solution is called the weak solution. It is obvious that the classical solutions of Eq.~(\ref{strong}) are also weak solutions. However, conversely, weak solutions must have sufficient smoothness to be the classical solutions, which dependents both on the shape of the boundary $\partial \Omega$ and the regularity of $c^2f$. For example, it can be shown that if $\Omega \subset \mathbb{R}^n$ has $C^k$ boundary (smooth up to $k$-th derivative) and $c^2f\in H^{k}(\Omega)$ with $k>\frac{n}{2}$, then $u\in C^2(\bar{\Omega})$, namely $u$ is a solution in the classical sense~\cite{nla:cat-vn1414651}. 

\subsection{Energy Conservation of the wave equation}
In addition to the {\it weak formulation} of the wave equation, another important property of the wave function Eq.~(\ref{strong}) is that it obeys the law of energy conservation. To see this point, we multiply Eq.~(\ref{strong}) by $\frac{\partial u}{\partial t}$ and integrate over the domain $\Omega$

\begin{align}
&\frac{1}{2}\frac{\partial}{\partial t}\int_{\Omega}\left[\left(\frac{\partial{u}}{\partial t}\right)^2+c^2\left(\nabla u\right)^2\right] \,\mathrm{d}x \nonumber \\
=& \int_{\partial\Omega}c^2\left(\frac{\partial u}{\partial t}\right)(\hat{n}\cdot\nabla u)\,\mathrm{d}x-\int_{\Omega}\left(\frac{\partial u}{\partial t}\right)\left[\left(\nabla c^2\right)\cdot\nabla u+c^2f\right]\,\mathrm{d}x \,. \label{Energy_con}
\end{align}

Note that the total energy carried by a wave in the domain $\Omega$ is defined by
\begin{equation}
E(t)=\frac{1}{2}\int_{\Omega}\left[\left(\frac{\partial{u}}{\partial t}\right)^2+c^2\left(\nabla u\right)^2\right] \,\mathrm{d}x \,. \label{defenergy}
\end{equation}
The left hand of the equality of Eq.~(\ref{Energy_con}) then represents the change of the total energy of wave with time.
From Eq.~(\ref{Energy_con}), if the boundary condition on the right hand side of the equality is a constant ($\frac{\partial{u}}{\partial t}|_{\partial \Omega}=0$) and there is no force source $f=0$ in the second term, for a constant speed of wave $\nabla c^2=0$, the total energy is conserved $\frac{dE(t)}{dt}=0$.

Equation~(\ref{Energy_con}) is one of the most important features of the wave function. We shall discuss this point in our following numerical analyses.

\section{The Finite Element Method \label{FEM}}
Before presenting our numerical analyses, we introduce an auxiliary function $v:=\frac{\partial u}{\partial t}$ and adopt the common notation 
\begin{equation}
\left(f,g\right)_{\Omega}=\int_{\Omega}f(x)g(x)\,\mathrm{d}x\quad,
\end{equation}
for convenience.
Equation~(\ref{weak}) in this case can be re-written as a set of two equations
\begin{equation}
 \left.\begin{aligned}
        &\left(\frac{\partial u}{\partial t},\phi\right)_{\Omega}-\left(v,\phi\right)_{\Omega}=0 \,,  \\
&-\left(\nabla u,\nabla(c^2\phi)\right)_{\Omega}-\left(\frac{\partial u}{\partial t},c\phi\right)_{\partial \Omega} \\
=&\left(\frac{\partial v}{\partial t},\phi \right)_{\Omega}+\left(c^2f,\phi\right)_{\Omega} 
       \end{aligned}
 \right \} \quad \forall \phi \in V \label{evtwo} \,.
\end{equation}
In order to solve the above equations numerically, we need to discretize these equations first. In general for a time-dependent problem, the discretization can be done either for the spatial variables first or the temporal variables first. The method to discretize the spatial variables first is called the (vertical) method
of lines (MOL). Once the spatial variables are discretized, each variable needs to be solved evolving with time. This constitutes a large system of ordinary differential equations (ODEs). The advantage of this method is that ODEs can be numerically solved by well-developed ODE solvers that are efficient and of high-order accuracy. The disadvantage of this method is that the spatial resolution has to be fixed at the very beginning. 

The alternative approach is called the {\it Rothe's method}, in which the temporal variable is discretized first. The advantage of this method is that the spatial resolution can be changed at each time step, which allows for the usage of the technique of {\it adaptive mesh refinement} (AMR). However, when trying to apply high-order temporal discretization schemes, {\it Rothe's method} is rather awkward.

Whether the time variable or the spatial variable should be discretized first has long been debated in the numerical analysis community. In this work, however, we follow the {\it Rothe's method} as we shall show that the spatial resolution is, in fact, the major limiting factor for the problem that we are interested in. The {\it Rothe's method} allows for great freedom to deal with the spatial resolution.

\subsection{discretization in time}
We use superscript $n$ to indicate the number of a time step and $k=t_{n}-t_{n-1}$ is the length of the present time step. We discretize the time derivative as 
\begin{equation}
\left \{
\begin{aligned}
\frac{\partial v}{\partial t}&\approx \frac{v^{n}-v^{n-1}}{k} \\
\frac{\partial u}{\partial t}&\approx \frac{u^{n}-u^{n-1}}{k}
\end{aligned}
\right. \,.
\end{equation}

The discretization scheme we adopted here also involves the spatial quantities at two different time steps, which is known as the $\theta$-scheme. Equation~(\ref{evtwo}) then can be represented as 
\begin{widetext}
\begin{equation}
\left. \begin{aligned}
\left(\frac{u^n-u^{n-1}}{k},\phi\right)_{\Omega}-\left(\theta v^n+(1-\theta)v^{n-1},\phi\right)_{\Omega} &=0 \\
-\left(\nabla [\theta u^n+(1-\theta)u^{n-1}],\nabla(c^2\phi)\right)_{\Omega}-\left(\frac{u^n-u^{n-1}}{k},c\phi\right)_{\partial \Omega} -\left(\frac{v^n-v^{n-1}}{k},\phi \right)_{\Omega}&=\left(c^2
[\theta f^n+(1-\theta)f^{n-1}],\phi\right)_{\Omega} 
\end{aligned} \right\}  \quad \forall \phi \in V \,. \label{Time_discretization}
\end{equation}
\end{widetext}

When $\theta = 0$, the scheme is called the forward or explicit Euler method. If $\theta = 0$, it reduces to the backward or implicit Euler method. The Euler method, no matter implicit or explicit, is only of first-order accurate. Therefore we do not adopt here.

The scheme we use here is called the {\it Crank-Nicolson Scheme}, namely $\theta = 1/2$, which uses the midpoint between two different time steps. This scheme is of second-order accuracy. As the midpoint rule is symplectic, the most important feature of this scheme is that it is energy-preserving. We shall discuss this point in our following numerical analyses.

\subsection{discretization in space}
We discretize the domain $\Omega \subset \mathbb{R}^3$ using the Finite Elements Methods (FEM). The FEM is an effective method for numerically solving partial differential equations. There are many excellent textbooks on this topic (Readers are referred to e.g. Ref.~\cite{FEMbook} for more details). Here, we will not go into too many mathematical details. Instead, we only summarise the main steps here.

First, the domain $\Omega$ is decomposed into subdomains $\Omega_i$, which consist of rectangles and triangles. This is called {\it decomposition} or {\it triangulation}. The vertices of rectangles and triangles in the domain $\Omega$ are called mesh points or nodes. Let $\Omega_h$ denote the set of all nodes of the decomposition. On each node, we construct a test function $\phi_i\in V\,,i=1,..,N$, where $V$ is the space defined in Eq.~(\ref{Vspace}) and $N$ is the total number of nodes in the domain. The test function $\phi_i$ is required to have the property
\begin{equation}
\phi_i(p^k)=\delta_{ik}, \quad i,k=1,..,N, \quad p^k\in \Omega_h \,, \nonumber
\end{equation}
where 
\begin{equation}
\delta_{ik} = \left \{
\begin{aligned}
1 \quad &{\rm for}& \quad i=k \\
0 \quad &{\rm for}& \quad i\ne k \nonumber
\end{aligned}
\right. \quad.
\end{equation}

Thus, $\phi_i$ has non-zero values only on the node with $k=i$ and its adjacent subdomains. It vanishes on other parts of the domain $\Omega$. The test function $\phi_i$ constructed in this way is called the {\it shape function}. Clearly, $\phi_i\in V$ on different nodes are linearly independent. We denote the space spanned by $\phi_i$ as
$V_h:=\mathrm{span}\{\phi_i\}_{i=1}^N$, which is a subspace of $V$. 

Next, we expand the scalar fields $u,v,c^2f$ in terms of $\phi_i$ 
\begin{equation}
u_h = \sum_{j=1}^N u_i \phi_i \,,
v_h = \sum_{j=1}^N v_i \phi_i \,,
c^2f_h = \sum_{j=1}^N c^2f_i \phi_i \,,\nonumber
\end{equation}
where $u_h\,,v_h\,,c^2f_h\in V_h$. These functions are approximations of the original scalar fields.
The advantage of this expansion is that the coefficients are precisely the values of these functions on the node points (e.g. $u_i=u_h(p_i)$).

Inserting the above expansions into Eq.~(\ref{Time_discretization}) and noting that Eq.~(\ref{Time_discretization}) holds for any $\phi\in V$, thus, we can choose $\phi$ as $\phi=\phi_j$, where $j=1,..,N$. Then we obtain $N$ different equations, which form a linear system
\begin{eqnarray} 
[k\theta(A+D)+B]U^n&=&-MV^n+G_2\nonumber\\
&-&k[\theta F^n+(1-\theta)F^{n-1}] \label{U1}\\
MU^n&=&k\theta MV^n+G_1 \label{V1}\,,
\end{eqnarray}
where $G_1$ and $G_2$ are defined by
\begin{equation}
\left \{
\begin{aligned} 
G_1&=MU^{n-1}+k(1-\theta)MV^{n-1}\\
G_2&=[-k(1-\theta)(A+D)+B]U^{n-1}+MV^{n-1}
\end{aligned}
\right.\,, \nonumber
\end{equation}
and the elements of the matrixes are defined by
\begin{equation}
\left \{
\begin{aligned} 
D_{ij}&=\left(\nabla \phi_i,\nabla (c^2) \phi_j\right)_{\Omega} \\
A_{ij}&=\left(\nabla \phi_i,c^2 \nabla \phi_j\right)_{\Omega}\\
M_{ij}&=\left(\phi_i,\phi_j\right)_{\Omega}\\
B_{ij}&=\left(c \phi_i, \phi_j\right)_{\partial \Omega}\\
F^n_{i}&=\left(c^2f^n, \phi_i\right)_{\Omega}\\
U^n_{i}&=\left(u^n, \phi_i\right)_{\Omega}\\
V^n_{i}&=\left(v^n, \phi_i\right)_{\Omega}
\end{aligned}
\right.\,. \nonumber
\end{equation}
Equations~(\ref{U1},\ref{V1}) are called the {\it Galerkin equations}, from which the unknown coefficients $u_i$
and $v_i$ can be solved. Note that Eq.~(\ref{U1}) explicitly contains $V^n$, which is unknown at the time step $n$. However, we can multiply Eq.~(\ref{U1}) by $k\theta$ and then use Eq.~(\ref{V1}) to eliminate $V^n$. Then we obtain 
\begin{widetext}
\begin{align}
[M+k^2\theta^2(A+D)+k\theta B]U^n&=G_1+k\theta G_2 - k^2\theta[\theta F^n+(1-\theta)F^{n-1}] \label{U2}\\
MV^n&=-[k \theta (A+D)+B]U^n+G_2 - k^2\theta[\theta F^n+(1-\theta)F^{n-1}] \label{V2}\quad.
\end{align}
\end{widetext}
Now, $U^n$ can be solved using the information on the time step $n-1$. Then $V^n$ can be solved with the knowledge of $U^n$. In practice, Eqs.~(\ref{U2},\ref{V2}) usually contain a very large number of equations. The {\it rank} of Eqs.~(\ref{U2},\ref{V2}) (the number of linear independent equations) is called the degrees-of-freedom (DOF) of the system, which depends on both the number of node points as well as the freedoms in the {\it shape functions}. A description of how to numerically solve these equations will be given in the next few subsections.

\subsection{Numerical Implement}
In this work, we use the public available code {\bf deal.ii}~\cite{dealII91,BangerthHartmannKanschat2007,dealII90} to numerically solve Eqs.~(\ref{U2},\ref{V2}). {\bf deal.ii} is a C++ program library aimed at numerically solving partial differential equations using modern finite element method. Coupled to stand-alone linear algebra libraries, such as PETSc~\cite{abhyankar2018petsc,petsc-web-page,petsc-user-ref,petsc-efficient},  {\bf deal.ii} supports massively parallel computing of vary large linear systems of equations. {\bf deal.ii} also provides convenient tools for {\it triangulation} of various geometries of the simulation domain. Therefore, it is straightforward to implement Eqs.~(\ref{U2},\ref{V2}) in the {\bf deal.ii} code. The detailed numerical analyses are presented in the next section.
 
\section{Code Tests\label{CT}} 
In this section, before going to simulate the scattering of GWs, we present several tests of our code using an important physical model that describes waves emitting from a point source. Moreover, we shall discuss the {\it Huygens-Fresnel principle} for wave functions. We use the {\it Huygens-Fresnel principle} to set boundary conditions so that we can avoid mathematical singularities of spherical waves at the origin.
\subsection{Spherical Waves from a Point Source and the Huygens-Fresnel principle}
The boundary-initial value problem for the propagation of waves from a point source can be presented as
\begin{eqnarray}
\nabla^2 u -\frac{1}{c^2}\frac{\partial^2}{\partial t^2}u =-Q(t)\delta(x) \quad {\rm in} \quad \Omega\times (0,T] \,, \label{pointsource}
\end{eqnarray}
where $Q(t)$ is the time-dependent waveform of the point source and $\delta$ is the Dirac delta function. The wave equation has a fundamental solution, which is the distributional solution of
\begin{equation}
\nabla^2 G -\frac{1}{c^2}\frac{\partial^2}{\partial t^2}G =-\delta(t)\delta(x) \,.
\end{equation}
The distribution $G$ is also called the Green's function. In free-space, the explicit form of $G$ for an outgoing wave is given by  
\begin{equation}
	G^{+}(x,t;x',t')=\frac{\delta(t'-[t-\frac{\left\|x-x'\right\|}{c}])}{4\pi\left\|x-x'\right\|}\,.
\end{equation}
The above equation is known as the {\it retarded Green's function} in free-space. The general solution of Eq.~(\ref{pointsource}) then has the form
\begin{align}
u(x,t) &= \int\int G^{+}(x,t;x',t')Q(t')\delta(x')dx'dt' \nonumber\\
       &= \frac{Q(t-\frac{r}{c})}{4\pi r} \,, \label{retarded_potential}
\end{align}
where $r$ is the norm of $x$, namely, $r=\left\|x\right\|$.

Although the wave function from a point source has a very simple analytical expression Eq.~(\ref{retarded_potential}), numerically solving Eq.~(\ref{pointsource}) is, indeed, non-trivial.
A challenge is that the point source on the right-hand side of Eq.~(\ref{pointsource}) is singular, which may cause numerical problems. In practice, in order to avoid this problem, we can make use of the Huygens-Fresnel principle to numerically solve Eq.~(\ref{pointsource}) on a slightly different domain without the origin but with new boundaries
\begin{align}
\nabla^2 u -\frac{1}{c^2}\frac{\partial^2}{\partial t^2}u &=0 \quad {\rm in} \quad \Omega/\{0\}\times (0,T]\,,
\end{align}
The Huygens-Fresnel principle states that the original waves from a source propagating to a distant observer can be considered as re-radiating from a wavefront of the original wave rather than directly from the source. As such, we can use the wavefront of the original wave at some radius as a new boundary for the wave equation. In this way, we can avoid the singularity in Eq.~(\ref{pointsource}) at the origin of the source. A mathematical description of the Huygens-Fresnel principle for spherical waves is provided in Appendix~\ref{Appen}. 

\begin{figure}
{\includegraphics[width=\linewidth]{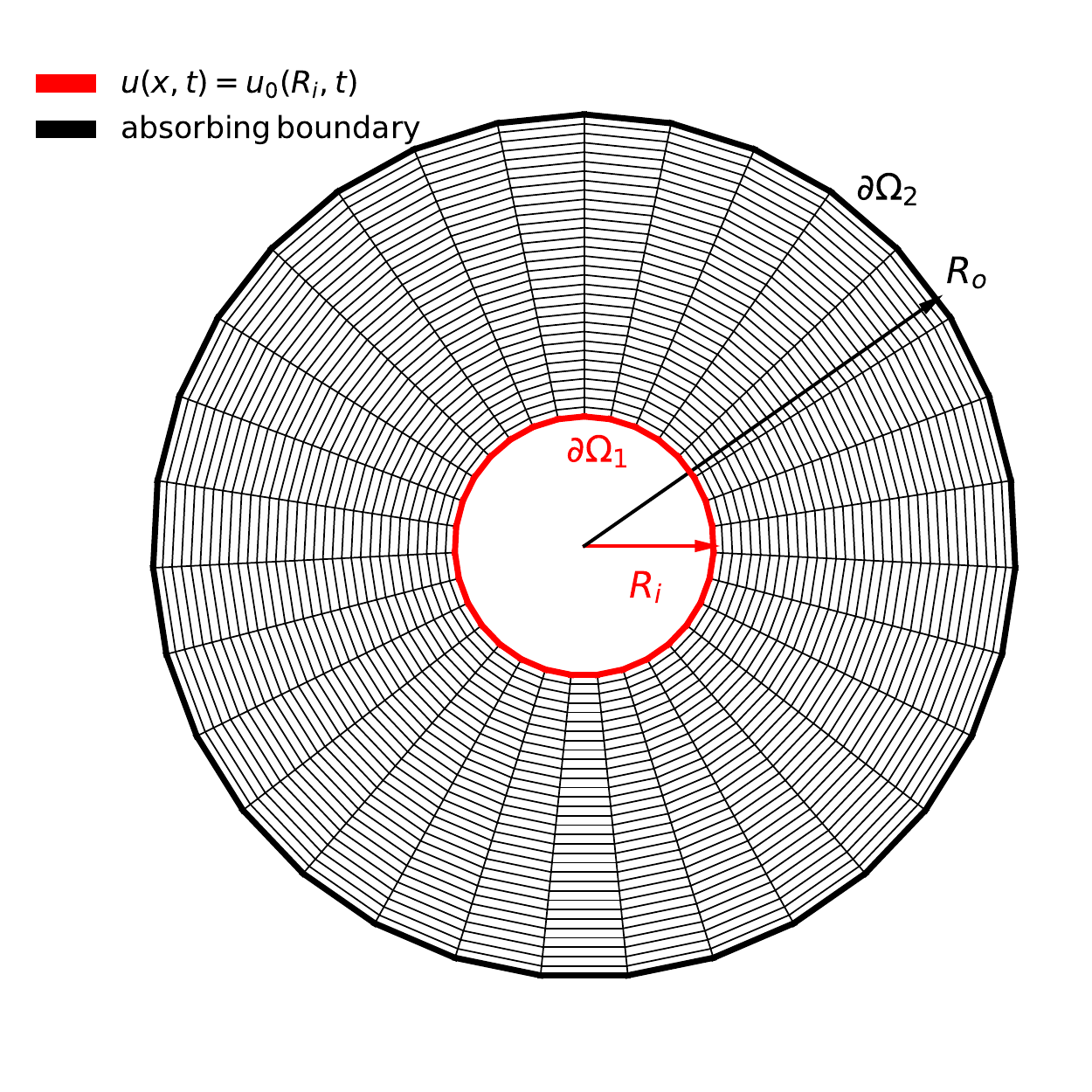}}
\caption{The boundaries and triangulation of our simulation domain. The domain consists of the bulk regions between the two spherical shells. The plot shows a slice taken through the center of the sphere. The meshes have a resolution of $2^5$. On $\partial \Omega_1$, we set the boundary conditions as the wavefront of the original wave from the point source propagating there. However, we set the absorbing boundary condition on $\partial \Omega_2$. \label{spherical_domain}}
\end{figure}

Figure~\ref{spherical_domain} shows the new boundaries and triangulation of our simulation domain. The domain consists of the bulk regions between the two spherical shells with an inner radius $R_i$ and an outer radius $R_o$, respectively. We set the boundary conditions at the inner shell $R_i$ as
\begin{equation}
u(R_i,t)=\left \{
	\begin{aligned} 
	&\frac{Q(t-\frac{R_i}{c})}{4\pi R_i} \quad &t\ge R_i/c\\
	&0\quad &t<R_i/c
	\end{aligned}
	\right.\quad {\rm on} \quad \partial \Omega_1\times (0,T]\,, \nonumber\\
\end{equation}
which is exactly the wavefront of the original point source propagating there. At the outer shell, we adopt the absorbing boundary condition
\begin{equation}
\frac{\partial u}{\partial r}(R_o,t) = - \frac{1}{c}\frac{\partial u}{\partial t} \quad {\rm on} \quad \partial \Omega_2\times (0,T]\,.
\end{equation}
The absorbing boundary condition is also called the non-reflecting boundary conditions or radiating boundary conditions. These boundary conditions can absorb and eliminate the reflections of waves on the boundaries. Therefore, with these boundary conditions, we can approximate the propagation of waves in free-space using a limited volume of the simulation domain. In addition to the boundary conditions, we set the initial conditions as
\begin{equation}
	u(x,0) = 0\quad {\rm in} \quad \Omega\,.
\end{equation}

After fixing the boundary and initial conditions, we perform several numerical tests. In these tests, we choose a concrete waveform $Q(t)=-A\sin(\omega t)$. Thus, the spherical wave has an analytical expression
\begin{equation}
u(x,t)=-\frac{A\sin[\omega (t-r/c)]}{4\pi r}\,, \label{waveanalytic}
\end{equation}
where $\omega = 2\pi f\,$ and $f$ is the frequency of waves. $A$ is the amplitude of the wave. In our test, we set $R_o=10^7M_{\odot}=49.2535{\rm Hz}^{-1}\,,$$R_i=9\times10^6M_{\odot}=44.32815{\rm Hz}^{-1}$ and $f=1$. Further, we set $A=4\pi R_i$, namely, the amplitude of the waves are normalized to unity at the inner boundary.

\subsection{Courant-Friedrichs-Levi condition and the time step}
To perform our simulations, we further need to set the time step. It is well known that the numerical methods for the time-dependent wave equation are only stable if the time step is small enough so that waves can have enough time to propagate through the space discretization. This condition is usually called the Courant-Friedrichs-Levi condition
\begin{equation}
	k\le\frac{\sigma}{c}\quad,
\end{equation}
where $\sigma$ is the size of the mesh and $k$ is the size of the time step as mentioned previously. But for a wave equation, the CFL condition alone can not guaranty a stable solution to the problem. This is because of the oscillating features of the waves. If the frequency of waves is too high, the wavelength will be too small and it is difficult to resolve the waveform within one period. The shape of the waves, in this case, is not smooth relative to the size of the meshes, which can lead to instabilities in the numerical simulations. Therefore, in order to get a stable solution, the time step and the size of the meshes should be small enough so that the waveform within one period can be well resolved. As such, in our simulations, we set $k=\lambda/30$ and the size of the mesh $\sigma<\lambda/7$. We have tested that further reducing the size of the mesh $\sigma$ does not improve the results appreciably.

\begin{figure}
{\includegraphics[width=\linewidth]{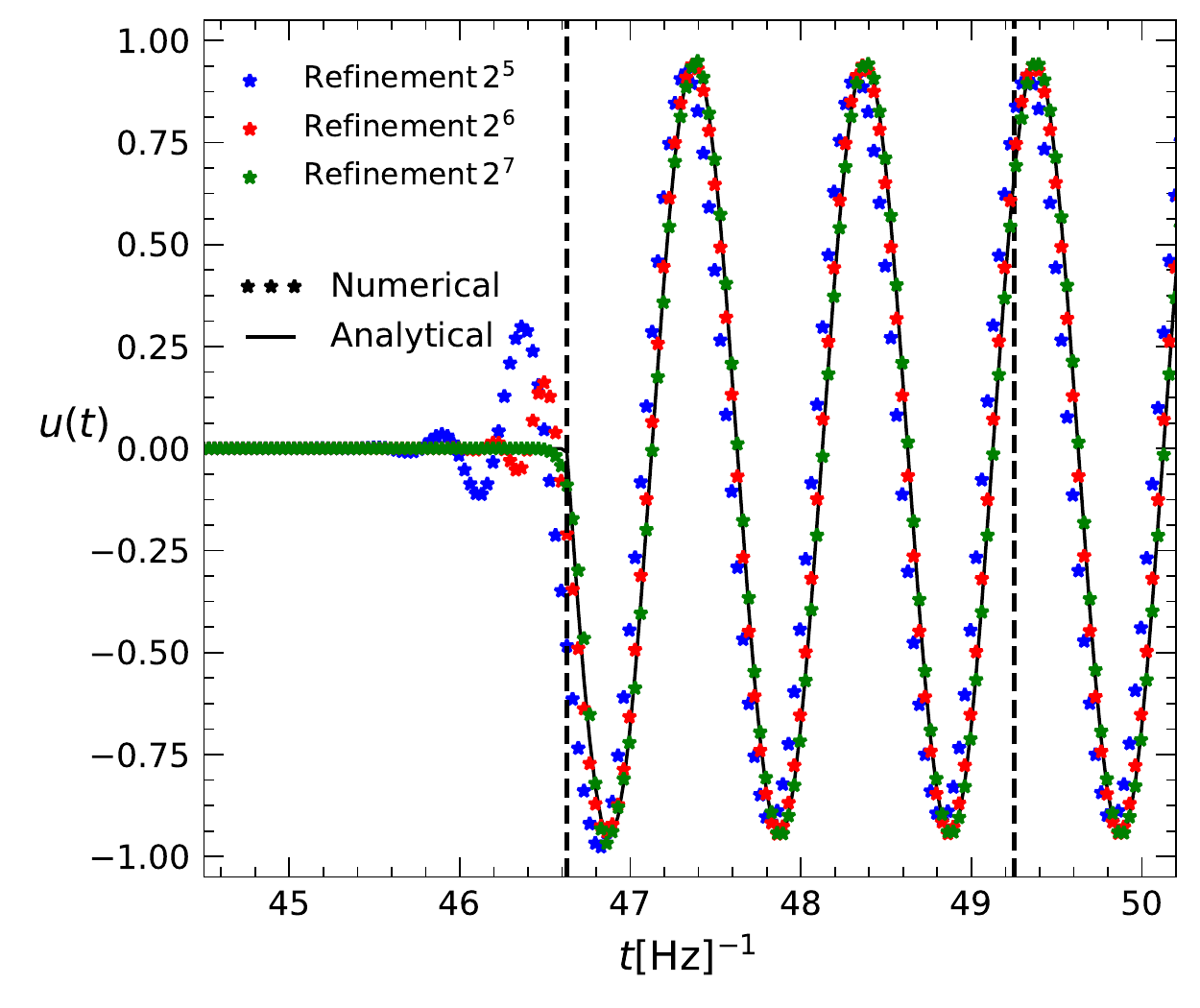}}
\caption{Numerical results (stars) against analytical results (solid lines).
The waveform is observed by an observer located at $r=46.63{\rm Hz}^{-1}$. The left vertical dashed line indicates the epoch that the wavefront just arrives at the observer. The right vertical dashed line indicates the epoch that the wavefront arrives at the outer boundary of our simulation domain. After this epoch, the wave starts getting out of the simulation domain. Symbols with different colors show the results obtained using different spatial resolutions.\label{SphericalWave}}
\end{figure}

\subsection{Numerical results}
We perform a suite of simulations with different spatial resolutions. In these simulations, we adopt the linear Lagrange test function, which is dependent only on the values of the finite elements on their vertices but does not involve any derivatives of the unknown fields. Therefore the degree of freedom (DOF) associated with each vertex is one. In this case, the total DOF is simply the total number of the expansion coefficients of the unknown field ($u_i$ or $v_i$). This number is also equal to the total number of linear algebraic equations in the system. The following table lists the level of refinement and the corresponding DOF.

\begin{table}[h!]
\begin{tabular}{c|c}
\hline 
Refinement & DOF \\ 
\hline 
$2^5$ & 202818 \\ 
\hline 
$2^6$ & 1597570 \\ 
\hline 
$2^7$ & 12681474 \\ 
\hline 
\end{tabular} 
\end{table}

Figure~\ref{SphericalWave} shows the waveform observed by an observer located at $r=46.63{\rm Hz}^{-1}$. Stars represent the numerical results and the solid lines are obtained from Eq.~(\ref{waveanalytic}).
The dashed vertical line indicates the epoch that the wavefront just arrives at the observer. Symbols with different colors show the results with different  resolutions. With high spatial resolution $2^7$ (blue stars), the numerical results agree with the analytic ones very well. This is expected as the accuracy of the FEM is strongly dependent on the resolution of elements.

After investigating the time-domain waveform for a particular local observer, we consider the global property of the wave equation. Inserting Eq.~(\ref{waveanalytic}) into Eq.~(\ref{defenergy}), we obtain the total energy of waves in the domain
\begin{align}
E(t)&=2\pi\omega^2R_i^2(t-R_i)\nonumber \\
+&\pi R_i\left\{\omega R_i \sin[2\omega(t-R_i)]-\cos[2\omega(t-R_i)]+1\right\}\,,\nonumber\\ 
& \quad t\in[R_i,R_o]\,. \label{totalenergy}
\end{align}
Then we measure the total energy in our simulation numerically. 

\begin{figure}
{\includegraphics[width=\linewidth]{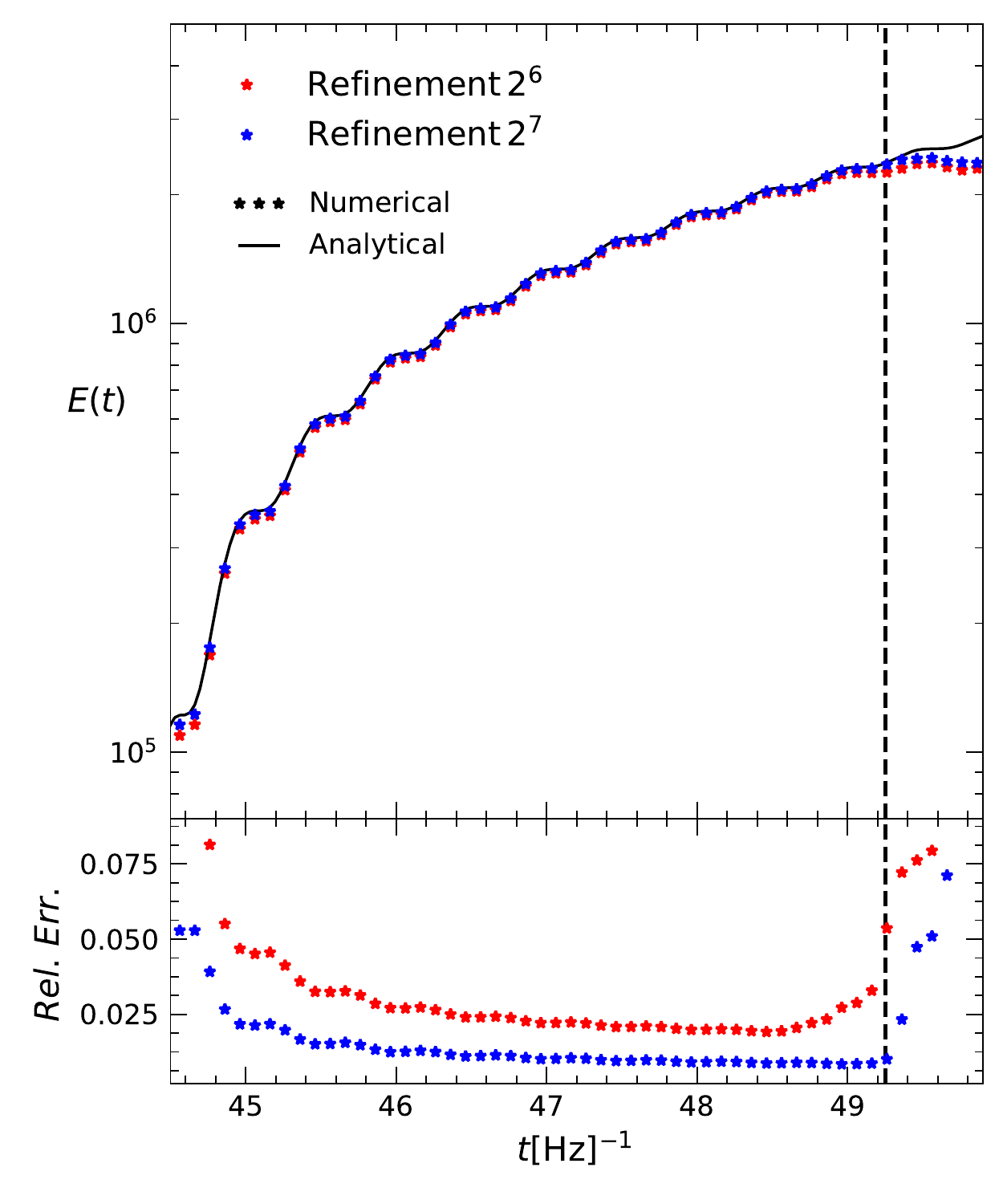}}
\caption{Upper panel: Comparisons of the total energy between numerical simulations (stars) and the exact analytic solution (solid line). Red and blue colors are for spatial resolution $2^6$ and $2^7$, respectively. The dashed vertical line indicates the epoch that the wavefront arrives at the outer boundary of our simulation domain. After $t>R_o$, waves start propagating out of the simulation domain. Lower Panel: relative errors between the numerical solutions and the exact analytic solution. In FEM, spatial resolution is the main driver of errors. By increasing the spatial resolution, the errors can be reduced significantly. \label{sphere_energy}}
\end{figure}

Figure~\ref{sphere_energy} compares the numerical results (stars) with the analytical ones (solid line) from Eq.~(\ref{totalenergy}). When the wavefront has not reached the outer boundary, namely, $R_i<t<R_o$, the numerical results agree with the analytic ones very well. After $t>R_o$ (vertical dashed line), the wavefront starts getting out of the simulation domain. The total energy inside the bulk of our simulation domain becomes steady (flat) as, in this case, the energy gettings into the simulation domain equals the energy getting out of the domain.

\section{Scattering of GWs by compact objects\label{GWP}}
After testing our numerical code, in this section, we aim to simulate the scattering of GWs by compact objects using the FEM. We first introduce the basic equations of GWs in non-uniform spacetime and then discuss the numerical methods for solving these equations. 
\subsection{basic equations}
We assume a metric $g_{\mu\nu}$ given by
\begin{equation}
ds^2=-(1+2\psi)dt^2+(1-2\psi)dx^2\,, \label{metric}
\end{equation}
and the gravitational waves as a perturbation to the background metric
\begin{equation}
	g_{\mu\nu}=g_{\mu\nu}^{0}+h_{\mu\nu}\quad,
\end{equation}
where $g_{\mu\nu}^{0}$ is the background metric. Following Ref.~\cite{Peters}, we neglect higher order non-linear terms and arrive at the propagation equation for gravitational waves $h_{ij}$
\begin{equation}
\nabla^2h_{ij}-(1-4\psi)\frac{\partial^2}{\partial t^2}h_{ij} = 0 \,.
\end{equation}

Using the eikonal approximation by Ref.~(\cite{Baraldo:1999ny}) (also see Ref.~\cite{GR_steven_weinberg}), the gravitational wave tensor can be represented as 
\begin{equation}
h_{ij}=h e_{ij}\,,
\end{equation}
where $e_{ij}$ is the polarization tensor. In this work, we assume that the polarization tensor does not change during the propagation of GW and then we obtain the so-called scalar wave equation
\begin{equation}
\nabla^2h-(1-4\psi)\frac{\partial^2h}{\partial t^2} = 0 \,. \label{Wave_potential}
\end{equation}

Equation~(\ref{Wave_potential}) is the core equation we aim to solve in this work. In principle, we could directly simulate the spherical waves from the origin to the scatterers given the potential $\psi$ of scatterers. However, if the scatterer is too distant from the origin, unlike in the previous section, it is difficult to simulate the spherical wave directly. This is because the radius of the sphere from the origin to the distant observer is too large. It is difficult to get enough spatial resolution for such a large sphere.  

Therefore, instead of simulating the full spherical waves, we only simulate the scattered waves near the scatterer. In order to do this, we first define the scattered wave as $\delta h := h-\tilde{h}$, where $\tilde{h}$ is the wave function that is unaffected by $\psi$, namely, the solution of Eq.~(\ref{Wave_potential}) with $\psi=0$. From the previous section, $\tilde{h}$ has an analytic expression
\begin{equation}
\tilde{h}(x,t)=\left \{
	\begin{aligned} 
	&\frac{Q(t-\frac{r}{c})}{4\pi r} \quad &t\ge r/c\\
	&0\quad &t<r/c
	\end{aligned}
	\right.
\end{equation}
where $Q(t)$ is the waveform of the GW source. Inserting the above equation and the expression $h=\delta h+\tilde{h}$ into Eq.~(\ref{Wave_potential}), we obtain
\begin{eqnarray}
c^2\nabla^2 \delta h -\frac{\partial^2}{\partial t^2}\delta h =-4c^2\psi \frac{\partial^2}{\partial t^2}\tilde{h} \,, \label{deltau}
\end{eqnarray}
where we have used
\begin{align}
\nabla^2 \tilde{h} -\frac{\partial^2}{\partial t^2}\tilde{h} &=0 \quad {\rm in} \quad \Omega/\{0\}\times (0,T]\,,
\end{align}
and $c^2=1/(1-4\psi)$ is the effective speed of wave. In free-space, using Green's function the above equation has a formal solution 
\begin{align}
\delta h =\frac{1}{\pi}\int dR\frac{c^2}{R}\psi \frac{\partial^2\delta h}{\partial t^2}\bigg\rvert_{t-\frac{R}{c}} \quad. \label{Greendeltau}	
\end{align}
Note that in Fourier-domain, the above equation is exactly the same as Eq.(3) in Ref.~\cite{Takahashi:2005sxa}. 

Given the fact that $\psi<0$, we have $c<1$. The presence of potential thus delays the propagation of GWs. Moreover, in order to make $\delta h(x,t)$ mathematically well defined, there must be some requirements on the smoothness of the source term $\psi$. In this work, we place a minimum requirement on the smoothness of $\psi$, namely,
\begin{equation}
	\psi \in H^1(\Omega)\,.
\end{equation}

\subsection{Plane wave approximation inside the simulation box}

\begin{figure}
{\includegraphics[width=\linewidth]{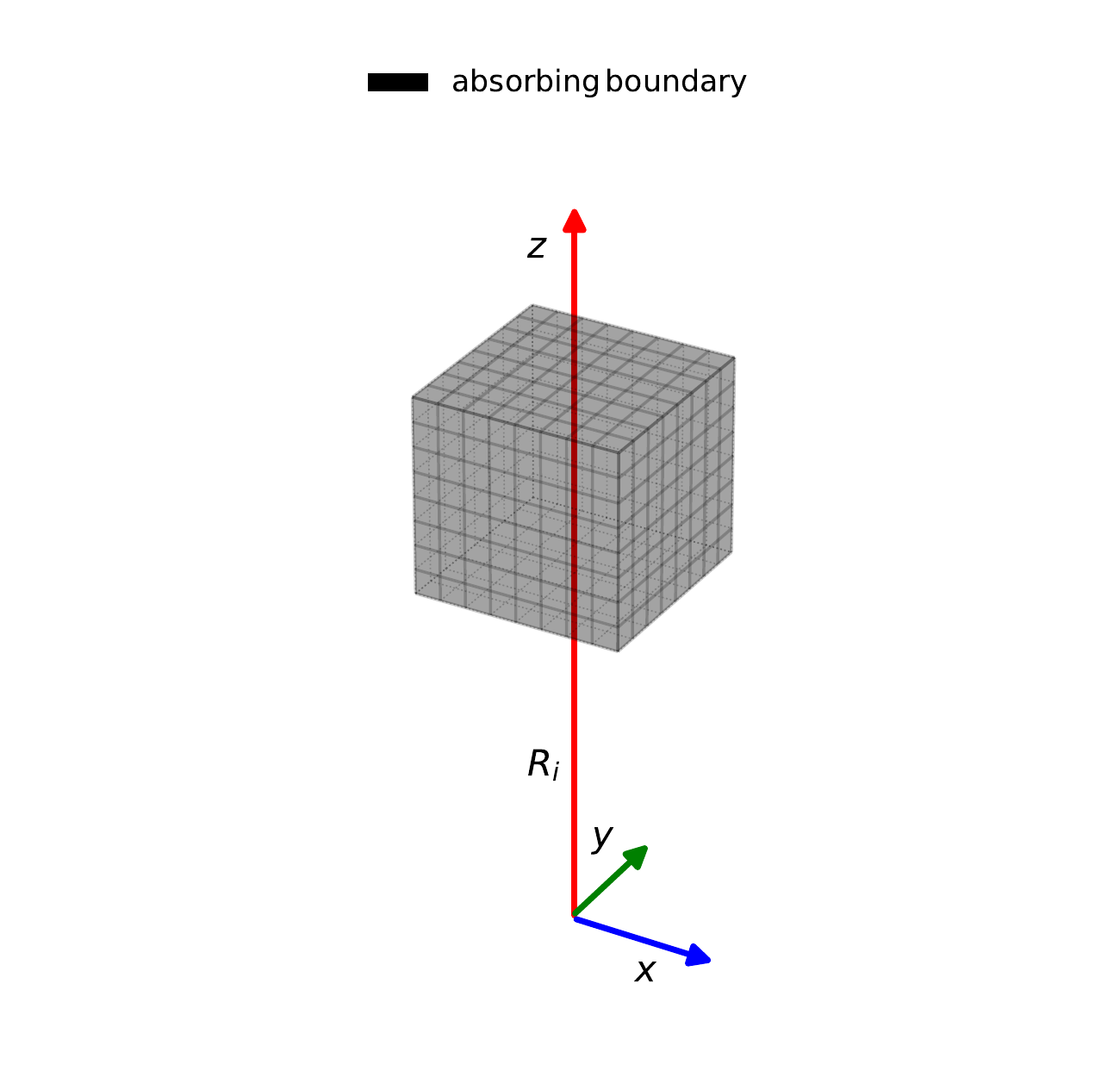}}
\caption{The domain of our simulations. We set the absorbing boundary conditions on all surfaces of the simulation box. $R_i$ is the distance to the source of the GWs. In our simulations, we choose $R_i = 20\,{\rm Mpc}$. As $R_i$ is very large, the incident spherical GWs can be locally treated as plane waves inside the simulation box. But for a distant observer far away from the simulation box, the incident GWs is still spherical. In this case, the scattered GWs can also be considered as spherical waves radiating from the scatterers at the center of the simulation box.    \label{Cubeboundary}}
\end{figure} 

As stated in the previous section, we only need to simulate the scattered waves around the scatterers. Figure~\ref{Cubeboundary} shows the domain and boundaries of our simulations. The simulation box is put along the $z$-axis, which is distant from the source of the GWs (e.g. $R_i = 20\,{\rm Mpc}$). The GWs inside the simulation box can be well approximated by plane waves. We set the initial condition as $$\delta h|_{z=R_i}=0\,,$$which means that the wavefront is simply the original wavefront at the initial epoch $h=\tilde{h}$. For other surfaces of the simulation box, we adopt the absorbing boundary conditions as noted before. The absorbing boundary conditions aim to guarantee that the scattered GWs $\delta h$ do not reflect into the simulation box when they arrive at the boundaries.

For $\psi$, we choose it as the potential generated by a homogeneous spherical ball located at $x_0=(0,0,R_i+L_{\rm box}/2)\,$, where $L_{\rm box}$ is the size of the simulation box along one side
\begin{equation}
\psi=\left \{
	\begin{aligned} 
	&-\frac{M}{\left\|x-x_0\right\|} \quad &\left\|x-x_0\right\|> R_s\\
	&-M\frac{3R_s^2-\left\|x-x_0\right\|^2}{2R_s^3}\quad &\left\|x-x_0\right\|\le R_s
	\end{aligned}
	\right. \,,
\end{equation}
where $R_s = 2M$ is the Schwarzschild radius. Note that, $\psi$ is continuous up to the first order of derivative, namely, $\psi\in C^1(\Omega)\subset H^1(\Omega)$. Further, we choose the waveform of the source of the GWs as $$Q(t)=-A\sin(\omega t)\,.$$ We set the frequency as $f=1$. The amplitude of the wave is normalized at $R_i$, namely, $A=4\pi R_i$.

\subsection{The effect of box size}
\begin{figure*}
{\includegraphics[width=\linewidth]{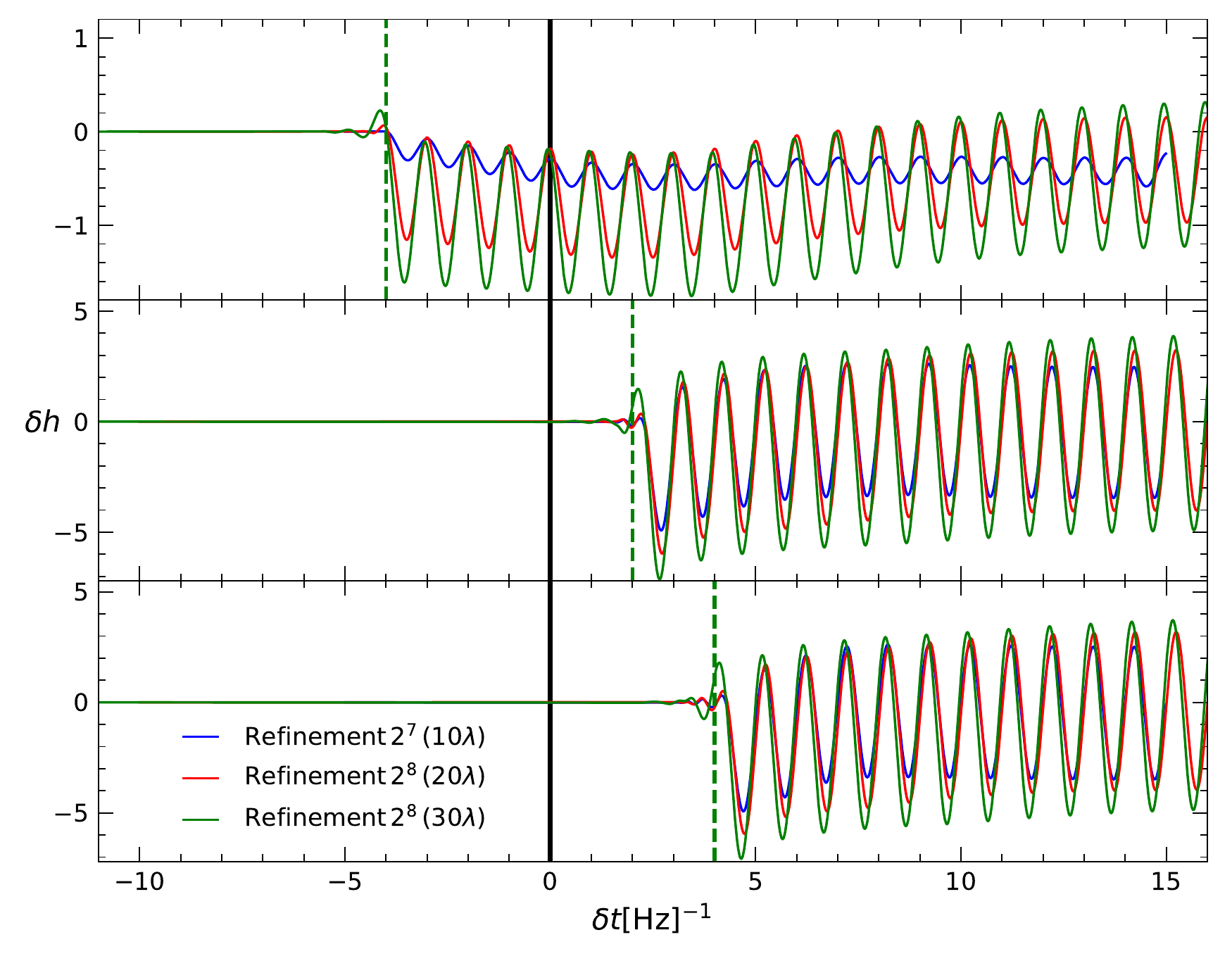}}
\caption{The scattered waves $\delta h$ as a function of time for different observers along the $z$-axis. Different colour are for simulations with different box sizes $L_{\rm box}=10\lambda\,,20\lambda\,,30\lambda\,$ respectively (blue,red, green lines). We set the absorbing boundary conditions at all the surfaces of the simulation box. As time and space have the same unit, the temporal axis can also be considered as the $z$-axis, along which the waves propagate. We also use $\delta t = t-(R_i+L_{\rm box}/2)$ as the temporal axis, for convenience, which is to set the zero point of time as the epoch when the wavefront arrives at the center of the simulation box. The top panel shows $\delta h$ for an observer located before the compact scatterer. In this case, $\delta h$ varies dramatically for simulations with different box sizes. Note that the simulations with $L_{\rm box}=10\lambda$ and $L_{\rm box}=20\lambda$ have the same temporal and spatial resolutions. Therefore, such differences between the blue and red lines are purely due to the changes in the size of the box. However, for observers located behind the compact object, due to the fact that most of the integral signals in Eq.~(\ref{Greendeltau}) come from the compact object, the box size has less significant effects on $\delta h$. \label{boxsize}}
\end{figure*}

After setting up the simulation system, we first investigate the effect of box size on our simulation results. To do this, in Figure~\ref{boxsize}, we present a suite of simulations with different box sizes $L_{\rm box}=10\lambda\,,20\lambda\,,30\lambda\,,$ where $\lambda=\frac{1}{f}$ is the wavelength. In these simulations, we choose the mass of the compact object as $M=10^4M_{\odot}=0.0492535{\rm Hz}^{-1}$. In this case, $\lambda$ is much greater than that of the Schwarzschild radius of the compact object. We choose local observers around the scatterers along the $z$-axis in the simulation box. We adopt $\delta t = t-(R_i+L_{\rm box}/2)$ as the temporal axis for convenience, which is to set the zero point of time as the epoch when the wavefront arrives at the center of the simulation box. The solid black vertical lines indicate the epoch that the wavefront of the GWs arrives at the location of the compact object (box center). We choose three observers along the $z$-axis inside the simulation box. The dashed vertical lines indicate GWs arriving at different observers. The top panel shows the scattered GWs $\delta h$ for an observer located before the scatterer. The middle and bottom ones show $\delta h$ for observers behind the compact object. Note that the simulations with $L_{\rm box}=10\lambda$ and $L_{\rm box}=20\lambda$ have the same temporal and spatial resolutions. So in these two simulations, the differences are purely due to the effects of the box size. It is clear that if the observers are before the compact object, as shown in the top panel, $\delta h$ varies dramatically for simulations with different box sizes. However, if the observers are behind the compact object, the effect of box size is limited. This is because, from Eq.~(\ref{Greendeltau}), the potential of the compact scatterer contributes most of the scattered signals for observers behind it. Further note that, the simulation with $L_{\rm box}=30\lambda\,$ has a lower spatial resolution. The diameter of the spatial finite element is about $\lambda/5$. There are clear spikes at the wavefront (bottom panel) which, however, is supposed to be zero. Therefore, part of the small differences in the middle and bottom panels between $L_{\rm box}=20\lambda\,$ and $L_{\rm box}=30\lambda\,$ runs are due to the insufficient spatial resolution in the $L_{\rm box}=30\lambda\,$ case rather than due to the effects of box size.

Indeed, it is difficult to eliminate the effect of box size. But such an effect can be significantly reduced by increasing both the resolution and volume of the simulation box. Arguably, it is only a matter of balance between scientific goals and computing resources. If we are only interested in signals behind the compact object, the current approach can provide reasonable accuracy on the scattered signal.

\subsection{Energy conservation}
\begin{figure}
{\includegraphics[width=\linewidth]{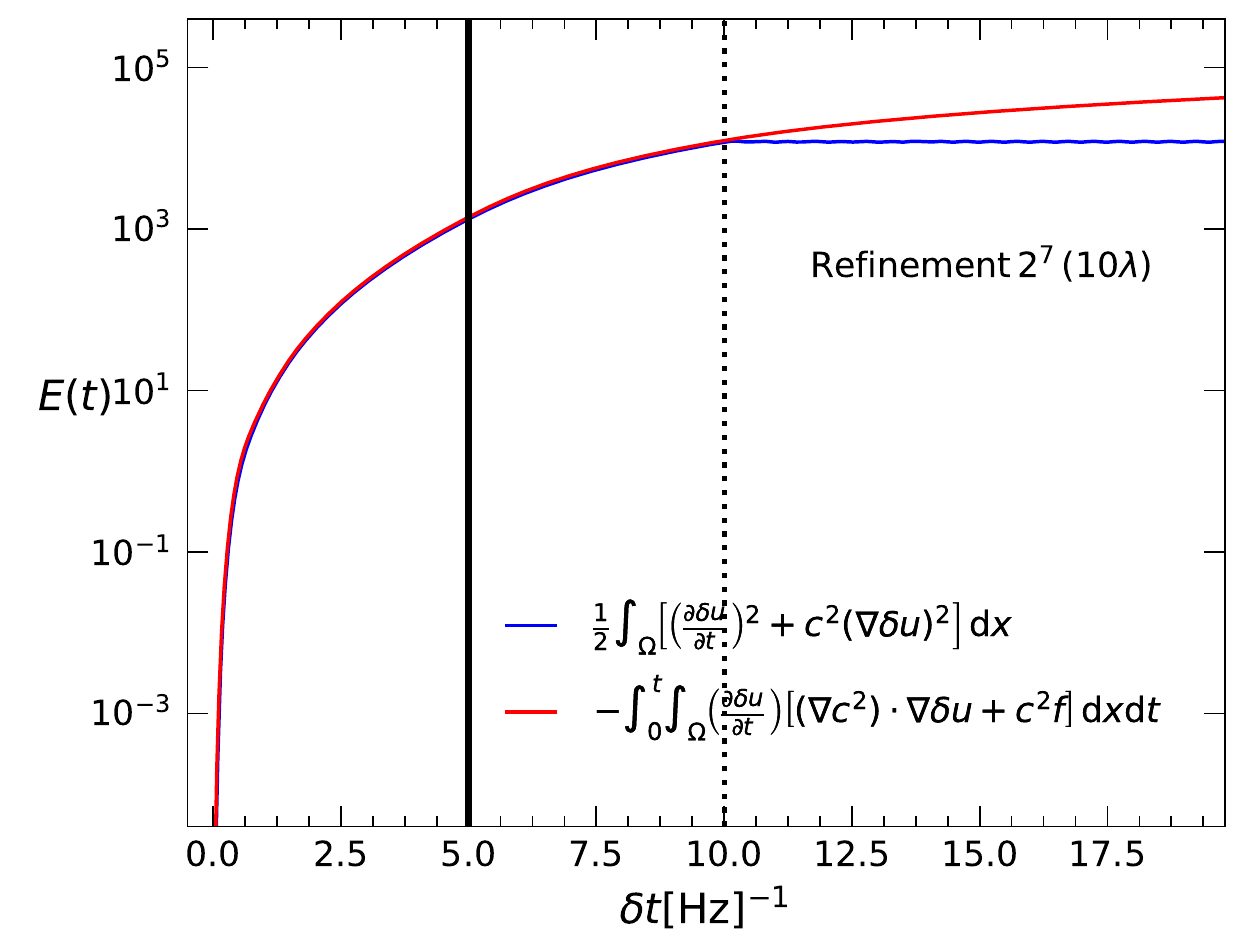}}
\caption{Energy conservation for the scattered wave $\delta h$. The blue lines represent the total energy measured in the simulation box. The red lines are the total energy generated by the source. The solid vertical line indicates the epoch that the wavefront of GWs arrives at the position of the compact object. The dashed vertical line indicates the epoch GWs arrive at the far-end boundary of the simulation box along the $z$-axis. Before the GWs passing through the whole simulation box, the total energy inside the box increases and equals to the energy generated from the source. The red and blue lines in this case overlap. However, when GWs get out of the box, the total energy inside the box becomes steady and conserved (blues lines after the dashed vertical line). In this case, the amount of energy generated by the source equals the amount of energy radiating out of the box.\label{Energy_num}}
\end{figure}

As already noted, one important property of the wave equation is that it obeys the law of energy conservation. This applies to $\delta h$ as well. To illustrate this point, in Fig.~\ref{Energy_num} we compare the numerical integration of the left-hand side and the right-hand side of Eq.~(\ref{Energy_con}). The size of the simulation box we choose in this test is $L_{\rm box}=10\lambda$. We adopt $\delta t = t-R_i$ as the temporal axis for convenience. The vertical black line indicates the epoch of the wavefront at the position of the compact object. The dashed line indicates the epoch when the GWs arrive at the far end of the surface of the simulation box along the $z$-axis.
When the wave passes through the whole simulation box, the total energy inside the simulation box becomes steady (blue lines). In this case, the amount of energy generated by the source equals the amount of energy radiating out of the box.

\subsection{Phase shifts nearby the scatterer}
We focus on $L_{\rm box}=20\lambda$ for the following discussions. The full GWs (incident+scattered) are given by
\begin{equation}
h= \delta h +\tilde{h} =\delta h -\frac{A\sin\left[\omega (t-\frac{r}{c})\right]}{4\pi r}\,.
\end{equation}
The wave effects of GWs are strongly dependent on the relative size between the wavelength and the Schwarzschild radius of the compact object. To show this point, we simulate the monochromatic waves with two different wavelengths $\lambda=1\,{\rm Hz}^{-1}$, $\lambda=2\,{\rm Hz}^{-1}$. 

\begin{figure*}
{\includegraphics[width=\linewidth]{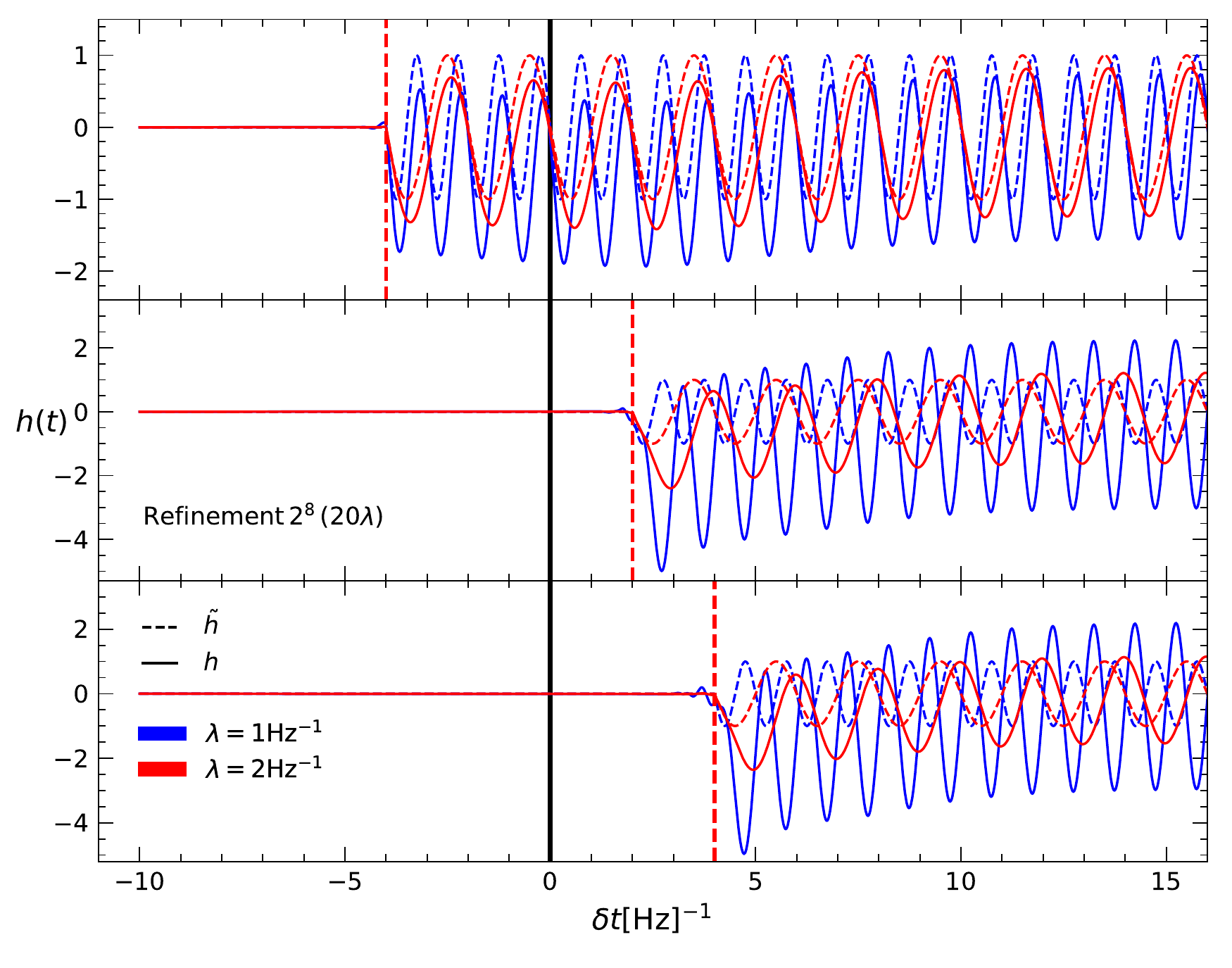}}
\caption{Numerical results for the total scattered (incident + scattered) monochromatic GWs with different wavelengths $\lambda=1\,{\rm Hz}^{-1}$ (blue), $\lambda=2\,{\rm Hz}^{-1}$ (red). The simulation has a box of $L_{\rm box}=20\lambda$ with a spatial resolution of $2^8$.  We use $\delta t = t-(R_i+L_{\rm box}/2)$ as the temporal axis for convenience. The dashed lines are for the unscattered signals and the solid lines are for the total scattered signals. The black solid vertical lines indicate the position of the compact object. The top panel shows $h$ for an observer located before the compact object. There are no significant changes in the waveform of the observed GWs. Behind the compact object, the lensing effect of the monochromatic wave with $\lambda=1\,{\rm Hz}^{-1}$ (blue lines) is very strong. There are significant changes in both the amplitude and the phases of waves. However, for the monochromatic wave with a much longer wavelength $\lambda=2\,{\rm Hz}^{-1}$ (red lines), the lensing effect is small due to the strong diffraction effect. In this case, the phases of the GWs are changed significantly.\label{Full_signal} }
\end{figure*}

Figure~\ref{Full_signal} shows the numerical results for three observers located along the $z$-axis before (top panel) and behind the compact object (middle and bottom panels). The dashed lines are for the unscattered signals and the solid lines are for the total scattered (incident+scattered) ones. Before the compact object, as expected, there are no significant changes in the waveform of the observed GWs neither in the amplitude nor the phase. Behind the compact object, both the amplitude and phase of GWs can be changed significantly. For a shorter wavelength $\lambda=1\,{\rm Hz}^{-1}$ (blue lines), the lensing effect of the monochromatic wave is very strong. The amplitude is increased by roughly a factor of two. However, for a much longer wavelength $\lambda=2\,{\rm Hz}^{-1}$ (red lines), the lensing effect and the changes in the amplitude are small due to diffraction. However, in this case, the phases of the original GWs are delayed significantly\footnote[1]{In textbooks of Quantum Mechanics, the phase shift is presented in terms of spatial stationary waves. For an attractive potential the phase shift is positive, namely, $e^{ik(r+a)}$, where $a$ is a positive constant. However, we focus on the time domain. The phase shift in time is negative $e^{ik(r+a)}e^{-i\omega t}=e^{ikr}e^{-i\omega (t-a)}$. }.

\subsection{Distant observer}
Indeed, from Eq.~(\ref{Greendeltau}) the scattered wave is mainly driven by the potential of the scatterer. Outside the scattering region, if the potential of the scatterer is negligible $\psi \sim 0$, from Eq.~(\ref{deltau}), the scattered GWs propagates in a free-space. In this case, the wave obeys
\begin{align}
c^2\nabla^2 \delta h -\frac{\partial^2}{\partial t^2}\delta h =0 \quad.
\end{align}
As shown in the Appendix, for a distant observer, the above equation has an analytical solution
\begin{align}
\delta h (R,\theta,\phi,t)=\frac{b}{R}\delta h_b(\theta,\phi,t-\frac{R-b}{c})\,, \label{farfield}
\end{align}
where $\delta h_b(R,\theta,\phi,t)$ is the Huygens-Fresnel wavefront at the radius of $b$. $b$ represents the radius above which the potential of the scatterer is negligible. $R$ is the distance from the scatterer to the observer as shown in Fig.~\ref{SphericalWave}. The total waves then are simply the superposition of the incident waves and the scattered waves
\begin{equation}
h = \frac{Q(t-r/c)}{4\pi r}+\frac{b}{R}\delta h_b[t-(R-b)/c]\quad,
\end{equation}
where $Q(t)$ is the waveform of the source.

Note that Eq.~(\ref{farfield}) applies not only to the isotropic case of the scattered waves but also works for the anisotropic one. The beam of the scattered waves can be highly concentrated along some directions. In these directions, the amplitude of the scattered waves can be much stronger than that of the incident waves. The amplitude of the scattered waves, in this case, is proportional to $\propto \frac{1}{R_i}\times\frac{1}{R}$, where $R_i$ is the distance from the source to the scatterer. This is consistent with the predictions of {\it Kirchoff's} theory (see Eq.(5) in Ref.~\cite{Takahashi:2005sxa}). However, if the frequency of the incident waves is very low, the scattered waves would be very isotropic due to the strong diffraction effect. The amplitude of the scattered waves is comparable to that of the incident waves. In this case, the incident waves cannot be neglected. The total scattered waves (incident+scattered) should be considered. We will further discuss this point in three cases:
  
\begin{itemize}
	\item $R$ is comparable to $b$ ($R\sim b$). The observer is close to the scatterer and within the potential of the scatterer (e.g. the scatterer is the central black hole of our galaxy~\cite{Ruffa_1999}). The amplitude of the scattered waves, in this case, is comparable to that of the incident wave. The phase shift effect is significant due to the interference between the incident and scattered waves.
	\item The observer is distant from the scatterer and $R$ is comparable to, $R_i$, the distance from the source to the scatterer. Due to the strong diffraction effect, the amplitude of the scattered waves is comparable to that of the incident wave at the scatterer, such as in our simulations. In this case, the scattered wave then decays as $\propto \frac{1}{R_i}\times\frac{1}{R}\sim \frac{1}{R^2}$, when it propagates to the distant observer. However, the original incident wave only decays as $\propto \frac{1}{r}\sim\frac{1}{2R}$. The scattered waves, thus, die out much faster than that of the original incident wave. At the observer, the effect of scattering by compact objects is negligible. This is consistent with our intuition that waves are not affected by obstacles that are much smaller than their wavelengths due to the effect of diffraction.
	\item If the source and the scatterer are very close to each other, the effect of scattering is significant. For a distant observer, the source and the scatterer can be considered as a single source, namely, $h \approx \frac{1}{r}\left[\frac{Q(t-r/c)}{4\pi}+b\delta h_b(t-r/c)\right]$.
\end{itemize} 
 
\section{conclusions \label{cl}}
The detection of gravitational waves in the low-frequency regime provides a powerful tool to probe the properties of supermassive back holes (SMBHs) throughout cosmic history, which has profound implications in gravitational physics as well as galaxy formation. However, due to the wave nature of GWs, wave effects such as diffraction or inference may hamper future GW experiments to accurately infer the source information from the observed signals. 

Given the complexity of the wave effects, numerical simulations provide a powerful tool to fully investigate these effects. 
Based on the public available code {\bf deal.ii} as well as stand-alone linear algebra libraries, such as PETSc, we have developed a code to simulate the wave effects of GWs using the finite element method (FEM). We have robustly tested our code using a monochromatic spherical wave from a point source. We find that not only for the waveform at individual observer but also for the global conservation of energy, our numerical results agree with the analytical predictions very well. 

After testing our code, we have studied the scattering of GWs by compact objects. For simplicity, we assume that the compact object is of spherical symmetry and the distribution of mass is uniform. As noted, the scattered wave is mainly driven by the potential of the scatterer (see Eq.~(\ref{Greendeltau})). Outside the potential well of the scatterer, the scattered waves travel in free-space and decay very rapidly. If the observer is far away from the scatterer, the scattered waves are negligible for incident waves with wavelengths much larger than the Schwarzschild radius of the scatterer. This phenomenon is well consistent with our intuition that waves are not affected by obstacles that are much smaller than their wavelengths due to diffraction. However, if the observer is very close to the scatterer, the scattered waves can be very large (driven by the potential of the scatterer). Based on simulations, we find that the amplitude of the scattered wave can be as large as that of the incident wave. In this case, due to the interference of the incident and scattered waves, phase shifts between the incident waves and the total scattered waves (incident + scattered waves) are significant. Since the wave equation of GWs propagating in a non-uniform background spacetime is similar to the well-known Schr{\"o}dinger's equation in Quantum mechanics, the effect of phase shifts near the scatterer is, indeed, expected.

Our findings also have practical implications. GWs signals may be scattered by compact objects from the source, on their way to detectors, such as the central black hole of our galaxy~\cite{Ruffa_1999}. Their waveform and phase may vary due to the scattering, which, indeed, poses a challenge in the GW data analysis. This is because a technique called matched filtering is usually adopted in digging out the underlying GW signals from a very noisy background. In this technique, the waveform, especially the phase, of the GW signals have to be predicted accurately. If there is a slight difference between the predicted templates and the true GW signals, the maximum signal-to-noise ratio of GW signals can not be achieved due to the effect of phase cancelation (See e.g.~\cite{Maggiore:1999vm} for a review). In this case, it is difficult to accurately extract the properties of the source of the GWs. Therefore, if GWs are scattered by compact objects, templates that account for such effects should be adopted in the data analysis.

Finally, although we have demonstrated the effectiveness of our code using a monochromatic wave, our code can be generalized to any waveform in the time-domain straightforwardly. In contrast to most of the previous work that focus only on the frequency-domain, the advantage of the time-domain is that it can be directly related to the waveform of GW signals.
However, in this work, we have only considered the propagation of GWs as scalar waves. It is reported recently that the polarization plane defined in geometric optics is smeared due to the diffraction effects~\cite{Cusin:2019rmt}. We shall investigate this phenomenon in our future work. Aslo, in a series of follow-up papers, we shall extend our work to non-static potentials such as, those of a Kerr black hole ~\cite{Baraldo:1999ny}, and binaries of compact objects. We shall also explore the effectiveness of using the propagation of GWs as a tool to test the theory of gravity~\cite{Chesler:2017khz,Belgacem:2019pkk}. 

\section*{Appendix\label{Appen}}
\subsection{Spherical Wave in free-space}
We adopt the following convention for the Fourier transform 
\begin{align}  
u(r,\theta,\phi,t)&=\int  u(r,\theta,\phi,\omega)e^{-i\omega t}\, \mathrm{d}t \quad,\nonumber\\
u(r,\theta,\phi,\omega)&=\frac{1}{2\pi}\int  u(r,\theta,\phi,t)e^{i\omega t}\, \mathrm{d}t \quad.\nonumber
\end{align}
In the frequency-domain, Eq.~(\ref{strong}) in free-space becomes
\begin{equation}
\left[\nabla^2+(\omega/c)^2\right] u = 0\quad.
\end{equation}

For a diverging spherical wave radiating from the origin, the general solution of the above equation has the form
\begin{equation}
	u(r,\theta,\phi,\omega) =\sum_{lm}a_{lm}(\omega)h_l^{(1)}\left(\frac{\omega}{c}r\right)Y_{lm}(\theta,\phi)\quad,\label{u_general}
\end{equation}
where $h_l^{(1)}(x)$ is the spherical Hankel functions of the first kind, which is the solution of the spherical Bessel equation satisfying the radiation boundary condition at infinity. $Y_{lm}(\theta,\phi)$ is the spherical harmonics. $a_{lm}(\omega)$ are unknown coefficients, which can be determined by the boundary conditions.

We assume that the boundary condition for a spherical wave is given at the surface of a sphere with a radius of $R$ 
 $u(r,\theta,\phi,\omega)|_{\partial \Omega}=f_R(\theta,\phi,t)$, where $f_R(\theta,\phi,t)$ is a known-function. In the frequency-domain, we have 
\begin{equation}  
f_R(\theta,\phi,\omega)=\frac{1}{2\pi}\int  f_R(\theta,\phi,t)e^{i\omega t}\, \mathrm{d}t\quad.
\end{equation}
We expand  $f_R(\theta,\phi,\omega)$ in terms of spherical harmonics
\begin{equation}  
f_R(\theta,\phi,\omega)=\sum_{lm}b_{lm}(w)Y_{lm}(\theta,\phi)\quad.
\end{equation}

Comparing the above equation with Eq.~(\ref{u_general}) at the boundary, we obtain
\begin{equation}
	a_{lm}(\omega) h_l^{(1)}\left(\frac{\omega}{c}R\right) = b_{lm}(\omega)\quad.
\end{equation}
Thus, we have
\begin{align}
u(r,\theta,\phi,t)&=\int  e^{-i\omega t} u(r,\theta,\phi,\omega) \, \mathrm{d}t \nonumber\\\
&=\int e^{-i\omega t} \sum_{lm} \frac{h_l^{(1)}(\frac{\omega}{c}r)}{h_l^{(1)}(\frac{\omega}{c}R)} b_{lm}(w)Y_{lm}(\theta,\phi)\, \mathrm{d}t \quad. \label{sphere_general}
\end{align}
The above equation demonstrates the Huygens-Fresnel principle, namely, the propagation of waves is completely determined by the surface of the wavefront. 

If the wavefront at the boundary is of spherical symmetry and is the same as the wavefront propagating there from the point source at the origin
\begin{equation}
f_R(\theta,\phi,t) = \frac{Q(t-\frac{R}{c})}{4\pi R}\quad,
\end{equation}
there will be only monopole $l=0$ left in the expansions of Eq.~(\ref{sphere_general}). Further noting that
\begin{equation}
h_0^{(1)}(x)= -\frac{i}{x}e^{ix}\quad,
\end{equation}
Eq.~(\ref{sphere_general}) then reduces to
\begin{align}
u(r,\theta,\phi,t)&=\frac{R}{r}\int e^{-i\omega (t-\frac{r-R}{c})}f_R(\theta,\phi,\omega) \, \mathrm{d}t \nonumber\\
&=\frac{R}{r}f_R(\theta,\phi,t-\frac{r-R}{c})\nonumber \\
&=\frac{Q(t-\frac{r}{c})}{4\pi r}\nonumber\quad,
\end{align}
which is the same as the wave propergating directly to the radius $r$ from the origin.

In general cases, if $x=\frac{\omega}{c}r\gg1$
\begin{equation}
	h_l^{(1)}(x)\sim \frac{1}{x}e^{ix}(-i)^{l+1} \nonumber\,,
\end{equation}
we have
\begin{align}
	\frac{h_l^{(1)}(\frac{\omega}{c}r)}{h_l^{(1)}(\frac{\omega}{c}R)}\sim \frac{R}{r}e^{i\frac{\omega}{c}(r-R)}\,.
\end{align}
The wave function reduces to
\begin{align}
u(r,\theta,\phi,t)&=\frac{R}{r}\int e^{-i\omega (t-\frac{r-R}{c})}f_R(\theta,\phi,\omega) \, \mathrm{d}t \nonumber \\
&=\frac{R}{r}f_R(\theta,\phi,t-\frac{r-R}{c})\,.
\end{align}
This indicates that the anisotropic feature in the wave function at the radius $R$ can be preserved to a distant observer during the propagation of waves, if the waves travel in a free-space. The amplitude of the wave decreases by a factor of $\frac{R}{r}$, which is the same as the isotropic wave, which is due to the conservation of energy of waves.
\section*{Acknowledgement} 
J.H.H. acknowledges support of Nanjing University and part of this work used the DiRAC@Durham facility managed by the Institute for Computational Cosmology on behalf of the STFC DiRAC HPC Facility (www.dirac.ac.uk). The equipment was funded by BEIS capital funding via STFC capital grants ST/K00042X/1, ST/P002293/1, ST/R002371/1 and ST/S002502/1, Durham University and STFC operations grant ST/R000832/1. DiRAC is part of the National e-Infrastructure.
\bibliography{myref}

\begin{thebibliography}{47}%
\makeatletter
\providecommand \@ifxundefined [1]{%
 \@ifx{#1\undefined}
}%
\providecommand \@ifnum [1]{%
 \ifnum #1\expandafter \@firstoftwo
 \else \expandafter \@secondoftwo
 \fi
}%
\providecommand \@ifx [1]{%
 \ifx #1\expandafter \@firstoftwo
 \else \expandafter \@secondoftwo
 \fi
}%
\providecommand \natexlab [1]{#1}%
\providecommand \enquote  [1]{``#1''}%
\providecommand \bibnamefont  [1]{#1}%
\providecommand \bibfnamefont [1]{#1}%
\providecommand \citenamefont [1]{#1}%
\providecommand \href@noop [0]{\@secondoftwo}%
\providecommand \href [0]{\begingroup \@sanitize@url \@href}%
\providecommand \@href[1]{\@@startlink{#1}\@@href}%
\providecommand \@@href[1]{\endgroup#1\@@endlink}%
\providecommand \@sanitize@url [0]{\catcode `\\12\catcode `\$12\catcode
  `\&12\catcode `\#12\catcode `\^12\catcode `\_12\catcode `\%12\relax}%
\providecommand \@@startlink[1]{}%
\providecommand \@@endlink[0]{}%
\providecommand \url  [0]{\begingroup\@sanitize@url \@url }%
\providecommand \@url [1]{\endgroup\@href {#1}{\urlprefix }}%
\providecommand \urlprefix  [0]{URL }%
\providecommand \Eprint [0]{\href }%
\providecommand \doibase [0]{http://dx.doi.org/}%
\providecommand \selectlanguage [0]{\@gobble}%
\providecommand \bibinfo  [0]{\@secondoftwo}%
\providecommand \bibfield  [0]{\@secondoftwo}%
\providecommand \translation [1]{[#1]}%
\providecommand \BibitemOpen [0]{}%
\providecommand \bibitemStop [0]{}%
\providecommand \bibitemNoStop [0]{.\EOS\space}%
\providecommand \EOS [0]{\spacefactor3000\relax}%
\providecommand \BibitemShut  [1]{\csname bibitem#1\endcsname}%
\let\auto@bib@innerbib\@empty
\bibitem [{\citenamefont {Abbott}\ \emph {et~al.}(2016)\citenamefont {Abbott}
  \emph {et~al.}}]{Abbott:2016blz}%
  \BibitemOpen
  \bibfield  {author} {\bibinfo {author} {\bibfnamefont {B.~P.}\ \bibnamefont
  {Abbott}} \emph {et~al.} (\bibinfo {collaboration} {LIGO Scientific,
  Virgo}),\ }\href {\doibase 10.1103/PhysRevLett.116.061102} {\bibfield
  {journal} {\bibinfo  {journal} {Phys. Rev. Lett.}\ }\textbf {\bibinfo
  {volume} {116}},\ \bibinfo {pages} {061102} (\bibinfo {year} {2016})},\
  \Eprint {http://arxiv.org/abs/1602.03837} {arXiv:1602.03837 [gr-qc]}
  \BibitemShut {NoStop}%
\bibitem [{\citenamefont {Punturo}\ \emph {et~al.}(2010)\citenamefont {Punturo}
  \emph {et~al.}}]{Einstein_tele}%
  \BibitemOpen
  \bibfield  {author} {\bibinfo {author} {\bibfnamefont {M.}~\bibnamefont
  {Punturo}} \emph {et~al.},\ }\href {\doibase 10.1088/0264-9381/27/19/194002}
  {\bibfield  {journal} {\bibinfo  {journal} {Classical and Quantum Gravity}\
  }\textbf {\bibinfo {volume} {27}},\ \bibinfo {pages} {194002} (\bibinfo
  {year} {2010})}\BibitemShut {NoStop}%
\bibitem [{\citenamefont {Dwyer}\ \emph {et~al.}(2015)\citenamefont {Dwyer},
  \citenamefont {Sigg}, \citenamefont {Ballmer}, \citenamefont {Barsotti},
  \citenamefont {Mavalvala},\ and\ \citenamefont {Evans}}]{LIGO40}%
  \BibitemOpen
  \bibfield  {author} {\bibinfo {author} {\bibfnamefont {S.}~\bibnamefont
  {Dwyer}}, \bibinfo {author} {\bibfnamefont {D.}~\bibnamefont {Sigg}},
  \bibinfo {author} {\bibfnamefont {S.~W.}\ \bibnamefont {Ballmer}}, \bibinfo
  {author} {\bibfnamefont {L.}~\bibnamefont {Barsotti}}, \bibinfo {author}
  {\bibfnamefont {N.}~\bibnamefont {Mavalvala}}, \ and\ \bibinfo {author}
  {\bibfnamefont {M.}~\bibnamefont {Evans}},\ }\href@noop {} {\bibfield
  {journal} {\bibinfo  {journal} {Phys. Rev. D}\ }\textbf {\bibinfo {volume}
  {91}},\ \bibinfo {pages} {082001} (\bibinfo {year} {2015})}\BibitemShut
  {NoStop}%
\bibitem [{\citenamefont {Seoane}\ \emph {et~al.}(2013)\citenamefont {Seoane}
  \emph {et~al.}}]{eLISA}%
  \BibitemOpen
  \bibfield  {author} {\bibinfo {author} {\bibfnamefont {P.~A.}\ \bibnamefont
  {Seoane}} \emph {et~al.} (\bibinfo {collaboration} {eLISA}),\ }\href@noop {}
  {\  (\bibinfo {year} {2013})},\ \Eprint {http://arxiv.org/abs/1305.5720}
  {arXiv:1305.5720 [astro-ph.CO]} \BibitemShut {NoStop}%
\bibitem [{\citenamefont {Sato}\ \emph {et~al.}(2009)\citenamefont {Sato} \emph
  {et~al.}}]{Sato_2009}%
  \BibitemOpen
  \bibfield  {author} {\bibinfo {author} {\bibfnamefont {S.}~\bibnamefont
  {Sato}} \emph {et~al.},\ }\href {\doibase 10.1088/1742-6596/154/1/012040}
  {\bibfield  {journal} {\bibinfo  {journal} {Journal of Physics: Conference
  Series}\ }\textbf {\bibinfo {volume} {154}},\ \bibinfo {pages} {012040}
  (\bibinfo {year} {2009})}\BibitemShut {NoStop}%
\bibitem [{\citenamefont {{Hobbs}}\ \emph {et~al.}(2010)\citenamefont
  {{Hobbs}}, \citenamefont {{Archibald}}, \citenamefont {{Arzoumanian}},
  \citenamefont {{Backer}}, \citenamefont {{Bailes}}, \citenamefont {{Bhat}},
  \citenamefont {{Burgay}}, \citenamefont {{Burke-Spolaor}}, \citenamefont
  {{Champion}}, \citenamefont {{Cognard}}, \citenamefont {{Coles}},
  \citenamefont {{Cordes}}, \citenamefont {{Demorest}}, \citenamefont
  {{Desvignes}}, \citenamefont {{Ferdman}}, \citenamefont {{Finn}},
  \citenamefont {{Freire}}, \citenamefont {{Gonzalez}}, \citenamefont
  {{Hessels}}, \citenamefont {{Hotan}}, \citenamefont {{Janssen}},
  \citenamefont {{Jenet}}, \citenamefont {{Jessner}}, \citenamefont {{Jordan}},
  \citenamefont {{Kaspi}}, \citenamefont {{Kramer}}, \citenamefont
  {{Kondratiev}}, \citenamefont {{Lazio}}, \citenamefont {{Lazaridis}},
  \citenamefont {{Lee}}, \citenamefont {{Levin}}, \citenamefont {{Lommen}},
  \citenamefont {{Lorimer}}, \citenamefont {{Lynch}}, \citenamefont {{Lyne}},
  \citenamefont {{Manchester}}, \citenamefont {{McLaughlin}}, \citenamefont
  {{Nice}}, \citenamefont {{Oslowski}}, \citenamefont {{Pilia}}, \citenamefont
  {{Possenti}}, \citenamefont {{Purver}}, \citenamefont {{Ransom}},
  \citenamefont {{Reynolds}}, \citenamefont {{Sanidas}}, \citenamefont
  {{Sarkissian}}, \citenamefont {{Sesana}}, \citenamefont {{Shannon}},
  \citenamefont {{Siemens}}, \citenamefont {{Stairs}}, \citenamefont
  {{Stappers}}, \citenamefont {{Stinebring}}, \citenamefont {{Theureau}},
  \citenamefont {{van Haasteren}}, \citenamefont {{van Straten}}, \citenamefont
  {{Verbiest}}, \citenamefont {{Yardley}},\ and\ \citenamefont
  {{You}}}]{2010CQGra..27h4013H}%
  \BibitemOpen
  \bibfield  {author} {\bibinfo {author} {\bibfnamefont {G.}~\bibnamefont
  {{Hobbs}}}, \bibinfo {author} {\bibfnamefont {A.}~\bibnamefont
  {{Archibald}}}, \bibinfo {author} {\bibfnamefont {Z.}~\bibnamefont
  {{Arzoumanian}}}, \bibinfo {author} {\bibfnamefont {D.}~\bibnamefont
  {{Backer}}}, \bibinfo {author} {\bibfnamefont {M.}~\bibnamefont {{Bailes}}},
  \bibinfo {author} {\bibfnamefont {N.~D.~R.}\ \bibnamefont {{Bhat}}}, \bibinfo
  {author} {\bibfnamefont {M.}~\bibnamefont {{Burgay}}}, \bibinfo {author}
  {\bibfnamefont {S.}~\bibnamefont {{Burke-Spolaor}}}, \bibinfo {author}
  {\bibfnamefont {D.}~\bibnamefont {{Champion}}}, \bibinfo {author}
  {\bibfnamefont {I.}~\bibnamefont {{Cognard}}}, \bibinfo {author}
  {\bibfnamefont {W.}~\bibnamefont {{Coles}}}, \bibinfo {author} {\bibfnamefont
  {J.}~\bibnamefont {{Cordes}}}, \bibinfo {author} {\bibfnamefont
  {P.}~\bibnamefont {{Demorest}}}, \bibinfo {author} {\bibfnamefont
  {G.}~\bibnamefont {{Desvignes}}}, \bibinfo {author} {\bibfnamefont {R.~D.}\
  \bibnamefont {{Ferdman}}}, \bibinfo {author} {\bibfnamefont {L.}~\bibnamefont
  {{Finn}}}, \bibinfo {author} {\bibfnamefont {P.}~\bibnamefont {{Freire}}},
  \bibinfo {author} {\bibfnamefont {M.}~\bibnamefont {{Gonzalez}}}, \bibinfo
  {author} {\bibfnamefont {J.}~\bibnamefont {{Hessels}}}, \bibinfo {author}
  {\bibfnamefont {A.}~\bibnamefont {{Hotan}}}, \bibinfo {author} {\bibfnamefont
  {G.}~\bibnamefont {{Janssen}}}, \bibinfo {author} {\bibfnamefont
  {F.}~\bibnamefont {{Jenet}}}, \bibinfo {author} {\bibfnamefont
  {A.}~\bibnamefont {{Jessner}}}, \bibinfo {author} {\bibfnamefont
  {C.}~\bibnamefont {{Jordan}}}, \bibinfo {author} {\bibfnamefont
  {V.}~\bibnamefont {{Kaspi}}}, \bibinfo {author} {\bibfnamefont
  {M.}~\bibnamefont {{Kramer}}}, \bibinfo {author} {\bibfnamefont
  {V.}~\bibnamefont {{Kondratiev}}}, \bibinfo {author} {\bibfnamefont
  {J.}~\bibnamefont {{Lazio}}}, \bibinfo {author} {\bibfnamefont
  {K.}~\bibnamefont {{Lazaridis}}}, \bibinfo {author} {\bibfnamefont {K.~J.}\
  \bibnamefont {{Lee}}}, \bibinfo {author} {\bibfnamefont {Y.}~\bibnamefont
  {{Levin}}}, \bibinfo {author} {\bibfnamefont {A.}~\bibnamefont {{Lommen}}},
  \bibinfo {author} {\bibfnamefont {D.}~\bibnamefont {{Lorimer}}}, \bibinfo
  {author} {\bibfnamefont {R.}~\bibnamefont {{Lynch}}}, \bibinfo {author}
  {\bibfnamefont {A.}~\bibnamefont {{Lyne}}}, \bibinfo {author} {\bibfnamefont
  {R.}~\bibnamefont {{Manchester}}}, \bibinfo {author} {\bibfnamefont
  {M.}~\bibnamefont {{McLaughlin}}}, \bibinfo {author} {\bibfnamefont
  {D.}~\bibnamefont {{Nice}}}, \bibinfo {author} {\bibfnamefont
  {S.}~\bibnamefont {{Oslowski}}}, \bibinfo {author} {\bibfnamefont
  {M.}~\bibnamefont {{Pilia}}}, \bibinfo {author} {\bibfnamefont
  {A.}~\bibnamefont {{Possenti}}}, \bibinfo {author} {\bibfnamefont
  {M.}~\bibnamefont {{Purver}}}, \bibinfo {author} {\bibfnamefont
  {S.}~\bibnamefont {{Ransom}}}, \bibinfo {author} {\bibfnamefont
  {J.}~\bibnamefont {{Reynolds}}}, \bibinfo {author} {\bibfnamefont
  {S.}~\bibnamefont {{Sanidas}}}, \bibinfo {author} {\bibfnamefont
  {J.}~\bibnamefont {{Sarkissian}}}, \bibinfo {author} {\bibfnamefont
  {A.}~\bibnamefont {{Sesana}}}, \bibinfo {author} {\bibfnamefont
  {R.}~\bibnamefont {{Shannon}}}, \bibinfo {author} {\bibfnamefont
  {X.}~\bibnamefont {{Siemens}}}, \bibinfo {author} {\bibfnamefont
  {I.}~\bibnamefont {{Stairs}}}, \bibinfo {author} {\bibfnamefont
  {B.}~\bibnamefont {{Stappers}}}, \bibinfo {author} {\bibfnamefont
  {D.}~\bibnamefont {{Stinebring}}}, \bibinfo {author} {\bibfnamefont
  {G.}~\bibnamefont {{Theureau}}}, \bibinfo {author} {\bibfnamefont
  {R.}~\bibnamefont {{van Haasteren}}}, \bibinfo {author} {\bibfnamefont
  {W.}~\bibnamefont {{van Straten}}}, \bibinfo {author} {\bibfnamefont
  {J.~P.~W.}\ \bibnamefont {{Verbiest}}}, \bibinfo {author} {\bibfnamefont
  {D.~R.~B.}\ \bibnamefont {{Yardley}}}, \ and\ \bibinfo {author}
  {\bibfnamefont {X.~P.}\ \bibnamefont {{You}}},\ }\href {\doibase
  10.1088/0264-9381/27/8/084013} {\bibfield  {journal} {\bibinfo  {journal}
  {Classical and Quantum Gravity}\ }\textbf {\bibinfo {volume} {27}},\ \bibinfo
  {eid} {084013} (\bibinfo {year} {2010})},\ \Eprint
  {http://arxiv.org/abs/0911.5206} {arXiv:0911.5206 [astro-ph.SR]} \BibitemShut
  {NoStop}%
\bibitem [{\citenamefont {Baibhav}\ \emph {et~al.}(2019)\citenamefont {Baibhav}
  \emph {et~al.}}]{Baibhav:2019rsa}%
  \BibitemOpen
  \bibfield  {author} {\bibinfo {author} {\bibfnamefont {V.}~\bibnamefont
  {Baibhav}} \emph {et~al.},\ }\href@noop {} {\  (\bibinfo {year} {2019})},\
  \Eprint {http://arxiv.org/abs/1908.11390} {arXiv:1908.11390 [astro-ph.HE]}
  \BibitemShut {NoStop}%
\bibitem [{\citenamefont {Bellovary}\ \emph {et~al.}(2016)\citenamefont
  {Bellovary}, \citenamefont {Mac~Low}, \citenamefont {McKernan},\ and\
  \citenamefont {Ford}}]{Bellovary:2015ifg}%
  \BibitemOpen
  \bibfield  {author} {\bibinfo {author} {\bibfnamefont {J.~M.}\ \bibnamefont
  {Bellovary}}, \bibinfo {author} {\bibfnamefont {M.-M.}\ \bibnamefont
  {Mac~Low}}, \bibinfo {author} {\bibfnamefont {B.}~\bibnamefont {McKernan}}, \
  and\ \bibinfo {author} {\bibfnamefont {K.~E.~S.}\ \bibnamefont {Ford}},\
  }\href {\doibase 10.3847/2041-8205/819/2/L17} {\bibfield  {journal} {\bibinfo
   {journal} {Astrophys. J.}\ }\textbf {\bibinfo {volume} {819}},\ \bibinfo
  {pages} {L17} (\bibinfo {year} {2016})},\ \Eprint
  {http://arxiv.org/abs/1511.00005} {arXiv:1511.00005 [astro-ph.GA]}
  \BibitemShut {NoStop}%
\bibitem [{\citenamefont {Bartos}\ \emph {et~al.}(2017)\citenamefont {Bartos},
  \citenamefont {Kocsis}, \citenamefont {Haiman},\ and\ \citenamefont
  {Márka}}]{Bartos:2016dgn}%
  \BibitemOpen
  \bibfield  {author} {\bibinfo {author} {\bibfnamefont {I.}~\bibnamefont
  {Bartos}}, \bibinfo {author} {\bibfnamefont {B.}~\bibnamefont {Kocsis}},
  \bibinfo {author} {\bibfnamefont {Z.}~\bibnamefont {Haiman}}, \ and\ \bibinfo
  {author} {\bibfnamefont {S.}~\bibnamefont {Márka}},\ }\href {\doibase
  10.3847/1538-4357/835/2/165} {\bibfield  {journal} {\bibinfo  {journal}
  {Astrophys. J.}\ }\textbf {\bibinfo {volume} {835}},\ \bibinfo {pages} {165}
  (\bibinfo {year} {2017})},\ \Eprint {http://arxiv.org/abs/1602.03831}
  {arXiv:1602.03831 [astro-ph.HE]} \BibitemShut {NoStop}%
\bibitem [{\citenamefont {Stone}\ \emph {et~al.}(2017)\citenamefont {Stone},
  \citenamefont {Metzger},\ and\ \citenamefont {Haiman}}]{Stone:2016wzz}%
  \BibitemOpen
  \bibfield  {author} {\bibinfo {author} {\bibfnamefont {N.~C.}\ \bibnamefont
  {Stone}}, \bibinfo {author} {\bibfnamefont {B.~D.}\ \bibnamefont {Metzger}},
  \ and\ \bibinfo {author} {\bibfnamefont {Z.}~\bibnamefont {Haiman}},\ }\href
  {\doibase 10.1093/mnras/stw2260} {\bibfield  {journal} {\bibinfo  {journal}
  {Mon. Not. Roy. Astron. Soc.}\ }\textbf {\bibinfo {volume} {464}},\ \bibinfo
  {pages} {946} (\bibinfo {year} {2017})},\ \Eprint
  {http://arxiv.org/abs/1602.04226} {arXiv:1602.04226 [astro-ph.GA]}
  \BibitemShut {NoStop}%
\bibitem [{\citenamefont {Haehnelt}\ and\ \citenamefont
  {Kauffmann}(2000)}]{Haehnelt:2000sx}%
  \BibitemOpen
  \bibfield  {author} {\bibinfo {author} {\bibfnamefont {M.~G.}\ \bibnamefont
  {Haehnelt}}\ and\ \bibinfo {author} {\bibfnamefont {G.}~\bibnamefont
  {Kauffmann}},\ }\href {\doibase 10.1046/j.1365-8711.2000.03989.x} {\bibfield
  {journal} {\bibinfo  {journal} {Mon. Not. Roy. Astron. Soc.}\ }\textbf
  {\bibinfo {volume} {318}},\ \bibinfo {pages} {L35} (\bibinfo {year}
  {2000})},\ \Eprint {http://arxiv.org/abs/astro-ph/0007369}
  {arXiv:astro-ph/0007369 [astro-ph]} \BibitemShut {NoStop}%
\bibitem [{\citenamefont {Ferrarese}\ and\ \citenamefont
  {Merritt}(2000)}]{Ferrarese:2000se}%
  \BibitemOpen
  \bibfield  {author} {\bibinfo {author} {\bibfnamefont {L.}~\bibnamefont
  {Ferrarese}}\ and\ \bibinfo {author} {\bibfnamefont {D.}~\bibnamefont
  {Merritt}},\ }\href {\doibase 10.1086/312838} {\bibfield  {journal} {\bibinfo
   {journal} {Astrophys. J.}\ }\textbf {\bibinfo {volume} {539}},\ \bibinfo
  {pages} {L9} (\bibinfo {year} {2000})},\ \Eprint
  {http://arxiv.org/abs/astro-ph/0006053} {arXiv:astro-ph/0006053 [astro-ph]}
  \BibitemShut {NoStop}%
\bibitem [{\citenamefont {{Bontz}}\ and\ \citenamefont
  {{Haugan}}(1981)}]{Bontz}%
  \BibitemOpen
  \bibfield  {author} {\bibinfo {author} {\bibfnamefont {R.~J.}\ \bibnamefont
  {{Bontz}}}\ and\ \bibinfo {author} {\bibfnamefont {M.~P.}\ \bibnamefont
  {{Haugan}}},\ }\href {\doibase 10.1007/BF00654034} {\bibfield  {journal}
  {\bibinfo  {journal} {apss}\ }\textbf {\bibinfo {volume} {78}},\ \bibinfo
  {pages} {199} (\bibinfo {year} {1981})}\BibitemShut {NoStop}%
\bibitem [{\citenamefont {Ohanian}(1974)}]{Ohanian:1974ys}%
  \BibitemOpen
  \bibfield  {author} {\bibinfo {author} {\bibfnamefont {H.~C.}\ \bibnamefont
  {Ohanian}},\ }\href {\doibase 10.1007/BF01810927} {\bibfield  {journal}
  {\bibinfo  {journal} {Int. J. Theor. Phys.}\ }\textbf {\bibinfo {volume}
  {9}},\ \bibinfo {pages} {425} (\bibinfo {year} {1974})}\BibitemShut {NoStop}%
\bibitem [{\citenamefont {Deguchi}\ and\ \citenamefont
  {Watson}(1986)}]{PhysRevD.34.1708}%
  \BibitemOpen
  \bibfield  {author} {\bibinfo {author} {\bibfnamefont {S.}~\bibnamefont
  {Deguchi}}\ and\ \bibinfo {author} {\bibfnamefont {W.~D.}\ \bibnamefont
  {Watson}},\ }\href {\doibase 10.1103/PhysRevD.34.1708} {\bibfield  {journal}
  {\bibinfo  {journal} {Phys. Rev. D}\ }\textbf {\bibinfo {volume} {34}},\
  \bibinfo {pages} {1708} (\bibinfo {year} {1986})}\BibitemShut {NoStop}%
\bibitem [{\citenamefont {Meena}\ and\ \citenamefont
  {Bagla}(2019)}]{Meena:2019ate}%
  \BibitemOpen
  \bibfield  {author} {\bibinfo {author} {\bibfnamefont {A.~K.}\ \bibnamefont
  {Meena}}\ and\ \bibinfo {author} {\bibfnamefont {J.~S.}\ \bibnamefont
  {Bagla}},\ }\href@noop {} {\  (\bibinfo {year} {2019})},\ \Eprint
  {http://arxiv.org/abs/1903.11809} {arXiv:1903.11809 [astro-ph.CO]}
  \BibitemShut {NoStop}%
\bibitem [{\citenamefont {{Deguchi}}\ and\ \citenamefont
  {{Watson}}(1986)}]{Deguchi}%
  \BibitemOpen
  \bibfield  {author} {\bibinfo {author} {\bibfnamefont {S.}~\bibnamefont
  {{Deguchi}}}\ and\ \bibinfo {author} {\bibfnamefont {W.~D.}\ \bibnamefont
  {{Watson}}},\ }\href {\doibase 10.1086/164389} {\bibfield  {journal}
  {\bibinfo  {journal} {Astrophys. J.}\ }\textbf {\bibinfo {volume} {307}},\
  \bibinfo {pages} {30} (\bibinfo {year} {1986})}\BibitemShut {NoStop}%
\bibitem [{\citenamefont {{Schneider}}\ \emph {et~al.}(1992)\citenamefont
  {{Schneider}}, \citenamefont {{Ehlers}},\ and\ \citenamefont
  {{Falco}}}]{Schneider}%
  \BibitemOpen
  \bibfield  {author} {\bibinfo {author} {\bibfnamefont {P.}~\bibnamefont
  {{Schneider}}}, \bibinfo {author} {\bibfnamefont {J.}~\bibnamefont
  {{Ehlers}}}, \ and\ \bibinfo {author} {\bibfnamefont {E.~E.}\ \bibnamefont
  {{Falco}}},\ }\href {\doibase 10.1007/978-3-662-03758-4} {\emph {\bibinfo
  {title} {{Gravitational Lenses}}}}\ (\bibinfo  {publisher} {Springer},\
  \bibinfo {year} {1992})\BibitemShut {NoStop}%
\bibitem [{\citenamefont {Ruffa}(1999)}]{Ruffa_1999}%
  \BibitemOpen
  \bibfield  {author} {\bibinfo {author} {\bibfnamefont {A.~A.}\ \bibnamefont
  {Ruffa}},\ }\href {\doibase 10.1086/312015} {\bibfield  {journal} {\bibinfo
  {journal} {The Astrophysical Journal}\ }\textbf {\bibinfo {volume} {517}},\
  \bibinfo {pages} {L31} (\bibinfo {year} {1999})}\BibitemShut {NoStop}%
\bibitem [{\citenamefont {De~Paolis}\ \emph {et~al.}(2002)\citenamefont
  {De~Paolis}, \citenamefont {Ingrosso}, \citenamefont {Nucita},\ and\
  \citenamefont {Qadir}}]{DePaolis:2002tw}%
  \BibitemOpen
  \bibfield  {author} {\bibinfo {author} {\bibfnamefont {F.}~\bibnamefont
  {De~Paolis}}, \bibinfo {author} {\bibfnamefont {G.}~\bibnamefont {Ingrosso}},
  \bibinfo {author} {\bibfnamefont {A.~A.}\ \bibnamefont {Nucita}}, \ and\
  \bibinfo {author} {\bibfnamefont {A.}~\bibnamefont {Qadir}},\ }\bibfield
  {booktitle} {\emph {\bibinfo {booktitle} {{Conference on Astronomical
  Telescopes and Instrumenation Waikoloa, Hawaii, August 22-28, 2002}}},\
  }\href {\doibase 10.1051/0004-6361:20021258} {\bibfield  {journal} {\bibinfo
  {journal} {Astron. Astrophys.}\ }\textbf {\bibinfo {volume} {394}},\ \bibinfo
  {pages} {749} (\bibinfo {year} {2002})},\ \Eprint
  {http://arxiv.org/abs/astro-ph/0209149} {arXiv:astro-ph/0209149 [astro-ph]}
  \BibitemShut {NoStop}%
\bibitem [{\citenamefont {Takahashi}\ and\ \citenamefont
  {Nakamura}(2003)}]{Takahashi:2003ix}%
  \BibitemOpen
  \bibfield  {author} {\bibinfo {author} {\bibfnamefont {R.}~\bibnamefont
  {Takahashi}}\ and\ \bibinfo {author} {\bibfnamefont {T.}~\bibnamefont
  {Nakamura}},\ }\href {\doibase 10.1086/377430} {\bibfield  {journal}
  {\bibinfo  {journal} {Astrophys. J.}\ }\textbf {\bibinfo {volume} {595}},\
  \bibinfo {pages} {1039} (\bibinfo {year} {2003})},\ \Eprint
  {http://arxiv.org/abs/astro-ph/0305055} {arXiv:astro-ph/0305055 [astro-ph]}
  \BibitemShut {NoStop}%
\bibitem [{\citenamefont {{Nakamura}}\ and\ \citenamefont
  {{Deguchi}}(1999)}]{1999PThPS.133..137N}%
  \BibitemOpen
  \bibfield  {author} {\bibinfo {author} {\bibfnamefont {T.~T.}\ \bibnamefont
  {{Nakamura}}}\ and\ \bibinfo {author} {\bibfnamefont {S.}~\bibnamefont
  {{Deguchi}}},\ }\href {\doibase 10.1143/PTPS.133.137} {\bibfield  {journal}
  {\bibinfo  {journal} {Progress of Theoretical Physics Supplement}\ }\textbf
  {\bibinfo {volume} {133}},\ \bibinfo {pages} {137} (\bibinfo {year}
  {1999})}\BibitemShut {NoStop}%
\bibitem [{\citenamefont {Suyama}\ \emph {et~al.}(2005)\citenamefont {Suyama},
  \citenamefont {Takahashi},\ and\ \citenamefont {Michikoshi}}]{Suyama:2005mx}%
  \BibitemOpen
  \bibfield  {author} {\bibinfo {author} {\bibfnamefont {T.}~\bibnamefont
  {Suyama}}, \bibinfo {author} {\bibfnamefont {R.}~\bibnamefont {Takahashi}}, \
  and\ \bibinfo {author} {\bibfnamefont {S.}~\bibnamefont {Michikoshi}},\
  }\href {\doibase 10.1103/PhysRevD.72.043001} {\bibfield  {journal} {\bibinfo
  {journal} {Phys. Rev.}\ }\textbf {\bibinfo {volume} {D72}},\ \bibinfo {pages}
  {043001} (\bibinfo {year} {2005})},\ \Eprint
  {http://arxiv.org/abs/astro-ph/0505023} {arXiv:astro-ph/0505023 [astro-ph]}
  \BibitemShut {NoStop}%
\bibitem [{\citenamefont {Christian}\ \emph {et~al.}(2018)\citenamefont
  {Christian}, \citenamefont {Vitale},\ and\ \citenamefont
  {Loeb}}]{Christian:2018vsi}%
  \BibitemOpen
  \bibfield  {author} {\bibinfo {author} {\bibfnamefont {P.}~\bibnamefont
  {Christian}}, \bibinfo {author} {\bibfnamefont {S.}~\bibnamefont {Vitale}}, \
  and\ \bibinfo {author} {\bibfnamefont {A.}~\bibnamefont {Loeb}},\ }\href
  {\doibase 10.1103/PhysRevD.98.103022} {\bibfield  {journal} {\bibinfo
  {journal} {Phys. Rev.}\ }\textbf {\bibinfo {volume} {D98}},\ \bibinfo {pages}
  {103022} (\bibinfo {year} {2018})},\ \Eprint
  {http://arxiv.org/abs/1802.02586} {arXiv:1802.02586 [astro-ph.HE]}
  \BibitemShut {NoStop}%
\bibitem [{\citenamefont {D'Orazio}\ and\ \citenamefont
  {Di~Stefano}(2019)}]{DOrazio:2019ens}%
  \BibitemOpen
  \bibfield  {author} {\bibinfo {author} {\bibfnamefont {D.~J.}\ \bibnamefont
  {D'Orazio}}\ and\ \bibinfo {author} {\bibfnamefont {R.}~\bibnamefont
  {Di~Stefano}},\ }\href@noop {} {\  (\bibinfo {year} {2019})},\ \Eprint
  {http://arxiv.org/abs/1906.11149} {arXiv:1906.11149 [astro-ph.HE]}
  \BibitemShut {NoStop}%
\bibitem [{\citenamefont {Zakharov}\ and\ \citenamefont
  {Baryshev}(2002)}]{Zakharov_2002}%
  \BibitemOpen
  \bibfield  {author} {\bibinfo {author} {\bibfnamefont {A.~F.}\ \bibnamefont
  {Zakharov}}\ and\ \bibinfo {author} {\bibfnamefont {Y.~V.}\ \bibnamefont
  {Baryshev}},\ }\href {\doibase 10.1088/0264-9381/19/7/319} {\bibfield
  {journal} {\bibinfo  {journal} {Classical and Quantum Gravity}\ }\textbf
  {\bibinfo {volume} {19}},\ \bibinfo {pages} {1361} (\bibinfo {year}
  {2002})}\BibitemShut {NoStop}%
\bibitem [{\citenamefont {{Barrar}}\ and\ \citenamefont
  {{Copson}}(1950)}]{Kirchoff}%
  \BibitemOpen
  \bibfield  {author} {\bibinfo {author} {\bibfnamefont {R.}~\bibnamefont
  {{Barrar}}}\ and\ \bibinfo {author} {\bibfnamefont {E.}~\bibnamefont
  {{Copson}}},\ }\href@noop {} {\emph {\bibinfo {title} {{The mathematical
  theory of Huygens' principle 2d ed}}}}\ (\bibinfo  {publisher} {Oxford},\
  \bibinfo {year} {1950})\BibitemShut {NoStop}%
\bibitem [{\citenamefont {Takahashi}\ \emph {et~al.}(2005)\citenamefont
  {Takahashi}, \citenamefont {Suyama},\ and\ \citenamefont
  {Michikoshi}}]{Takahashi:2005sxa}%
  \BibitemOpen
  \bibfield  {author} {\bibinfo {author} {\bibfnamefont {R.}~\bibnamefont
  {Takahashi}}, \bibinfo {author} {\bibfnamefont {T.}~\bibnamefont {Suyama}}, \
  and\ \bibinfo {author} {\bibfnamefont {S.}~\bibnamefont {Michikoshi}},\
  }\href {\doibase 10.1051/0004-6361:200500140} {\bibfield  {journal} {\bibinfo
   {journal} {Astron. Astrophys.}\ }\textbf {\bibinfo {volume} {438}},\
  \bibinfo {pages} {L5} (\bibinfo {year} {2005})},\ \Eprint
  {http://arxiv.org/abs/astro-ph/0503343} {arXiv:astro-ph/0503343 [astro-ph]}
  \BibitemShut {NoStop}%
\bibitem [{\citenamefont {{Weinberg}}(1972)}]{GR_steven_weinberg}%
  \BibitemOpen
  \bibfield  {author} {\bibinfo {author} {\bibfnamefont {S.}~\bibnamefont
  {{Weinberg}}},\ }\href@noop {} {\emph {\bibinfo {title} {{Gravitation and
  cosmology: principles and applications of the general theory of
  relativity}}}}\ (\bibinfo  {publisher} {John Wilry and Sons, Inc.},\ \bibinfo
  {year} {1972})\BibitemShut {NoStop}%
\bibitem [{\citenamefont {Peters}(1974)}]{Peters}%
  \BibitemOpen
  \bibfield  {author} {\bibinfo {author} {\bibfnamefont {P.~C.}\ \bibnamefont
  {Peters}},\ }\href {\doibase 10.1103/PhysRevD.9.2207} {\bibfield  {journal}
  {\bibinfo  {journal} {Phys. Rev. D}\ }\textbf {\bibinfo {volume} {9}},\
  \bibinfo {pages} {2207} (\bibinfo {year} {1974})}\BibitemShut {NoStop}%
\bibitem [{\citenamefont {Sorge}(2015)}]{Sorge:2015yoa}%
  \BibitemOpen
  \bibfield  {author} {\bibinfo {author} {\bibfnamefont {F.}~\bibnamefont
  {Sorge}},\ }\href {\doibase 10.1088/0264-9381/32/3/035007} {\bibfield
  {journal} {\bibinfo  {journal} {Class. Quant. Grav.}\ }\textbf {\bibinfo
  {volume} {32}},\ \bibinfo {pages} {035007} (\bibinfo {year}
  {2015})}\BibitemShut {NoStop}%
\bibitem [{\citenamefont {Allen}(2005)}]{Allen:2004gu}%
  \BibitemOpen
  \bibfield  {author} {\bibinfo {author} {\bibfnamefont {B.}~\bibnamefont
  {Allen}},\ }\href {\doibase 10.1103/PhysRevD.71.062001} {\bibfield  {journal}
  {\bibinfo  {journal} {Phys. Rev.}\ }\textbf {\bibinfo {volume} {D71}},\
  \bibinfo {pages} {062001} (\bibinfo {year} {2005})},\ \Eprint
  {http://arxiv.org/abs/gr-qc/0405045} {arXiv:gr-qc/0405045 [gr-qc]}
  \BibitemShut {NoStop}%
\bibitem [{\citenamefont {Maggiore}(2000)}]{Maggiore:1999vm}%
  \BibitemOpen
  \bibfield  {author} {\bibinfo {author} {\bibfnamefont {M.}~\bibnamefont
  {Maggiore}},\ }\href {\doibase 10.1016/S0370-1573(99)00102-7} {\bibfield
  {journal} {\bibinfo  {journal} {Phys. Rept.}\ }\textbf {\bibinfo {volume}
  {331}},\ \bibinfo {pages} {283} (\bibinfo {year} {2000})},\ \Eprint
  {http://arxiv.org/abs/gr-qc/9909001} {arXiv:gr-qc/9909001 [gr-qc]}
  \BibitemShut {NoStop}%
\bibitem [{\citenamefont {{Zienkiewicz}}\ and\ \citenamefont
  {{Taylor}}(1989)}]{FEMbook}%
  \BibitemOpen
  \bibfield  {author} {\bibinfo {author} {\bibfnamefont {O.}~\bibnamefont
  {{Zienkiewicz}}}\ and\ \bibinfo {author} {\bibfnamefont {R.}~\bibnamefont
  {{Taylor}}},\ }\href@noop {} {\emph {\bibinfo {title} {{The finite element
  method. Vol.1, Basic formulation and linear problems}}}}\ (\bibinfo
  {publisher} {London:McGraw-Hill},\ \bibinfo {year} {1989})\BibitemShut
  {NoStop}%
\bibitem [{\citenamefont {Lax}\ and\ \citenamefont {Milgram}(1954)}]{LaxMilg}%
  \BibitemOpen
  \bibfield  {author} {\bibinfo {author} {\bibfnamefont {P.~D.}\ \bibnamefont
  {Lax}}\ and\ \bibinfo {author} {\bibfnamefont {A.~N.}\ \bibnamefont
  {Milgram}},\ }in\ \href@noop {} {\emph {\bibinfo {booktitle} {Contributions
  to the theory of partial differential equations}}},\ \bibinfo {series and
  number} {Annals of Mathematics Studies, no. 33}\ (\bibinfo  {publisher}
  {Princeton University Press, Princeton, N. J.},\ \bibinfo {year} {1954})\
  pp.\ \bibinfo {pages} {167--190}\BibitemShut {NoStop}%
\bibitem [{\citenamefont {{Grisvard}}(1985)}]{nla:cat-vn1414651}%
  \BibitemOpen
  \bibfield  {author} {\bibinfo {author} {\bibfnamefont {P.}~\bibnamefont
  {{Grisvard}}},\ }\href@noop {} {\emph {\bibinfo {title} {{ Elliptic problems
  in nonsmooth domains}}}}\ (\bibinfo  {publisher} {Pitman Advanced Pub.
  Program Boston},\ \bibinfo {year} {1985})\BibitemShut {NoStop}%
\bibitem [{\citenamefont {Arndt}\ \emph {et~al.}(2019)\citenamefont {Arndt},
  \citenamefont {Bangerth}, \citenamefont {Clevenger}, \citenamefont {Davydov},
  \citenamefont {Fehling}, \citenamefont {Garcia-Sanchez}, \citenamefont
  {Harper}, \citenamefont {Heister}, \citenamefont {Heltai}, \citenamefont
  {Kronbichler}, \citenamefont {Kynch}, \citenamefont {Maier}, \citenamefont
  {Pelteret}, \citenamefont {Turcksin},\ and\ \citenamefont
  {Wells}}]{dealII91}%
  \BibitemOpen
  \bibfield  {author} {\bibinfo {author} {\bibfnamefont {D.}~\bibnamefont
  {Arndt}}, \bibinfo {author} {\bibfnamefont {W.}~\bibnamefont {Bangerth}},
  \bibinfo {author} {\bibfnamefont {T.~C.}\ \bibnamefont {Clevenger}}, \bibinfo
  {author} {\bibfnamefont {D.}~\bibnamefont {Davydov}}, \bibinfo {author}
  {\bibfnamefont {M.}~\bibnamefont {Fehling}}, \bibinfo {author} {\bibfnamefont
  {D.}~\bibnamefont {Garcia-Sanchez}}, \bibinfo {author} {\bibfnamefont
  {G.}~\bibnamefont {Harper}}, \bibinfo {author} {\bibfnamefont
  {T.}~\bibnamefont {Heister}}, \bibinfo {author} {\bibfnamefont
  {L.}~\bibnamefont {Heltai}}, \bibinfo {author} {\bibfnamefont
  {M.}~\bibnamefont {Kronbichler}}, \bibinfo {author} {\bibfnamefont {R.~M.}\
  \bibnamefont {Kynch}}, \bibinfo {author} {\bibfnamefont {M.}~\bibnamefont
  {Maier}}, \bibinfo {author} {\bibfnamefont {J.-P.}\ \bibnamefont {Pelteret}},
  \bibinfo {author} {\bibfnamefont {B.}~\bibnamefont {Turcksin}}, \ and\
  \bibinfo {author} {\bibfnamefont {D.}~\bibnamefont {Wells}},\ }\href
  {\doibase 10.1515/jnma-2019-0064} {\bibfield  {journal} {\bibinfo  {journal}
  {Journal of Numerical Mathematics}\ } (\bibinfo {year} {2019}),\
  10.1515/jnma-2019-0064},\ \bibinfo {note} {accepted}\BibitemShut {NoStop}%
\bibitem [{\citenamefont {Bangerth}\ \emph {et~al.}(2007)\citenamefont
  {Bangerth}, \citenamefont {Hartmann},\ and\ \citenamefont
  {Kanschat}}]{BangerthHartmannKanschat2007}%
  \BibitemOpen
  \bibfield  {author} {\bibinfo {author} {\bibfnamefont {W.}~\bibnamefont
  {Bangerth}}, \bibinfo {author} {\bibfnamefont {R.}~\bibnamefont {Hartmann}},
  \ and\ \bibinfo {author} {\bibfnamefont {G.}~\bibnamefont {Kanschat}},\
  }\href@noop {} {\bibfield  {journal} {\bibinfo  {journal} {ACM Trans. Math.
  Softw.}\ }\textbf {\bibinfo {volume} {33}},\ \bibinfo {pages} {24/1}
  (\bibinfo {year} {2007})}\BibitemShut {NoStop}%
\bibitem [{\citenamefont {Alzetta}\ \emph {et~al.}(2018)\citenamefont
  {Alzetta}, \citenamefont {Arndt}, \citenamefont {Bangerth}, \citenamefont
  {Boddu}, \citenamefont {Brands}, \citenamefont {Davydov}, \citenamefont
  {Gassmoeller}, \citenamefont {Heister}, \citenamefont {Heltai}, \citenamefont
  {Kormann}, \citenamefont {Kronbichler}, \citenamefont {Maier}, \citenamefont
  {Pelteret}, \citenamefont {Turcksin},\ and\ \citenamefont
  {Wells}}]{dealII90}%
  \BibitemOpen
  \bibfield  {author} {\bibinfo {author} {\bibfnamefont {G.}~\bibnamefont
  {Alzetta}}, \bibinfo {author} {\bibfnamefont {D.}~\bibnamefont {Arndt}},
  \bibinfo {author} {\bibfnamefont {W.}~\bibnamefont {Bangerth}}, \bibinfo
  {author} {\bibfnamefont {V.}~\bibnamefont {Boddu}}, \bibinfo {author}
  {\bibfnamefont {B.}~\bibnamefont {Brands}}, \bibinfo {author} {\bibfnamefont
  {D.}~\bibnamefont {Davydov}}, \bibinfo {author} {\bibfnamefont
  {R.}~\bibnamefont {Gassmoeller}}, \bibinfo {author} {\bibfnamefont
  {T.}~\bibnamefont {Heister}}, \bibinfo {author} {\bibfnamefont
  {L.}~\bibnamefont {Heltai}}, \bibinfo {author} {\bibfnamefont
  {K.}~\bibnamefont {Kormann}}, \bibinfo {author} {\bibfnamefont
  {M.}~\bibnamefont {Kronbichler}}, \bibinfo {author} {\bibfnamefont
  {M.}~\bibnamefont {Maier}}, \bibinfo {author} {\bibfnamefont {J.-P.}\
  \bibnamefont {Pelteret}}, \bibinfo {author} {\bibfnamefont {B.}~\bibnamefont
  {Turcksin}}, \ and\ \bibinfo {author} {\bibfnamefont {D.}~\bibnamefont
  {Wells}},\ }\href {\doibase 10.1515/jnma-2018-0054} {\bibfield  {journal}
  {\bibinfo  {journal} {Journal of Numerical Mathematics}\ }\textbf {\bibinfo
  {volume} {26}},\ \bibinfo {pages} {173} (\bibinfo {year} {2018})}\BibitemShut
  {NoStop}%
\bibitem [{\citenamefont {Abhyankar}\ \emph {et~al.}(2018)\citenamefont
  {Abhyankar}, \citenamefont {Brown}, \citenamefont {Constantinescu},
  \citenamefont {Ghosh}, \citenamefont {Smith},\ and\ \citenamefont
  {Zhang}}]{abhyankar2018petsc}%
  \BibitemOpen
  \bibfield  {author} {\bibinfo {author} {\bibfnamefont {S.}~\bibnamefont
  {Abhyankar}}, \bibinfo {author} {\bibfnamefont {J.}~\bibnamefont {Brown}},
  \bibinfo {author} {\bibfnamefont {E.~M.}\ \bibnamefont {Constantinescu}},
  \bibinfo {author} {\bibfnamefont {D.}~\bibnamefont {Ghosh}}, \bibinfo
  {author} {\bibfnamefont {B.~F.}\ \bibnamefont {Smith}}, \ and\ \bibinfo
  {author} {\bibfnamefont {H.}~\bibnamefont {Zhang}},\ }\href@noop {}
  {\bibfield  {journal} {\bibinfo  {journal} {arXiv preprint arXiv:1806.01437}\
  } (\bibinfo {year} {2018})}\BibitemShut {NoStop}%
\bibitem [{\citenamefont {Balay}\ \emph
  {et~al.}(2019{\natexlab{a}})\citenamefont {Balay}, \citenamefont {Abhyankar},
  \citenamefont {Adams}, \citenamefont {Brown}, \citenamefont {Brune},
  \citenamefont {Buschelman}, \citenamefont {Dalcin}, \citenamefont {Dener},
  \citenamefont {Eijkhout}, \citenamefont {Gropp}, \citenamefont {Karpeyev},
  \citenamefont {Kaushik}, \citenamefont {Knepley}, \citenamefont {May},
  \citenamefont {McInnes}, \citenamefont {Mills}, \citenamefont {Munson},
  \citenamefont {Rupp}, \citenamefont {Sanan}, \citenamefont {Smith},
  \citenamefont {Zampini}, \citenamefont {Zhang},\ and\ \citenamefont
  {Zhang}}]{petsc-web-page}%
  \BibitemOpen
  \bibfield  {author} {\bibinfo {author} {\bibfnamefont {S.}~\bibnamefont
  {Balay}}, \bibinfo {author} {\bibfnamefont {S.}~\bibnamefont {Abhyankar}},
  \bibinfo {author} {\bibfnamefont {M.~F.}\ \bibnamefont {Adams}}, \bibinfo
  {author} {\bibfnamefont {J.}~\bibnamefont {Brown}}, \bibinfo {author}
  {\bibfnamefont {P.}~\bibnamefont {Brune}}, \bibinfo {author} {\bibfnamefont
  {K.}~\bibnamefont {Buschelman}}, \bibinfo {author} {\bibfnamefont
  {L.}~\bibnamefont {Dalcin}}, \bibinfo {author} {\bibfnamefont
  {A.}~\bibnamefont {Dener}}, \bibinfo {author} {\bibfnamefont
  {V.}~\bibnamefont {Eijkhout}}, \bibinfo {author} {\bibfnamefont {W.~D.}\
  \bibnamefont {Gropp}}, \bibinfo {author} {\bibfnamefont {D.}~\bibnamefont
  {Karpeyev}}, \bibinfo {author} {\bibfnamefont {D.}~\bibnamefont {Kaushik}},
  \bibinfo {author} {\bibfnamefont {M.~G.}\ \bibnamefont {Knepley}}, \bibinfo
  {author} {\bibfnamefont {D.~A.}\ \bibnamefont {May}}, \bibinfo {author}
  {\bibfnamefont {L.~C.}\ \bibnamefont {McInnes}}, \bibinfo {author}
  {\bibfnamefont {R.~T.}\ \bibnamefont {Mills}}, \bibinfo {author}
  {\bibfnamefont {T.}~\bibnamefont {Munson}}, \bibinfo {author} {\bibfnamefont
  {K.}~\bibnamefont {Rupp}}, \bibinfo {author} {\bibfnamefont {P.}~\bibnamefont
  {Sanan}}, \bibinfo {author} {\bibfnamefont {B.~F.}\ \bibnamefont {Smith}},
  \bibinfo {author} {\bibfnamefont {S.}~\bibnamefont {Zampini}}, \bibinfo
  {author} {\bibfnamefont {H.}~\bibnamefont {Zhang}}, \ and\ \bibinfo {author}
  {\bibfnamefont {H.}~\bibnamefont {Zhang}},\ }\href
  {https://www.mcs.anl.gov/petsc} {\enquote {\bibinfo {title} {{PETS}c {W}eb
  page},}\ }\bibinfo {howpublished} {\url{https://www.mcs.anl.gov/petsc}}
  (\bibinfo {year} {2019}{\natexlab{a}})\BibitemShut {NoStop}%
\bibitem [{\citenamefont {Balay}\ \emph
  {et~al.}(2019{\natexlab{b}})\citenamefont {Balay}, \citenamefont {Abhyankar},
  \citenamefont {Adams}, \citenamefont {Brown}, \citenamefont {Brune},
  \citenamefont {Buschelman}, \citenamefont {Dalcin}, \citenamefont {Dener},
  \citenamefont {Eijkhout}, \citenamefont {Gropp}, \citenamefont {Karpeyev},
  \citenamefont {Kaushik}, \citenamefont {Knepley}, \citenamefont {May},
  \citenamefont {McInnes}, \citenamefont {Mills}, \citenamefont {Munson},
  \citenamefont {Rupp}, \citenamefont {Sanan}, \citenamefont {Smith},
  \citenamefont {Zampini}, \citenamefont {Zhang},\ and\ \citenamefont
  {Zhang}}]{petsc-user-ref}%
  \BibitemOpen
  \bibfield  {author} {\bibinfo {author} {\bibfnamefont {S.}~\bibnamefont
  {Balay}}, \bibinfo {author} {\bibfnamefont {S.}~\bibnamefont {Abhyankar}},
  \bibinfo {author} {\bibfnamefont {M.~F.}\ \bibnamefont {Adams}}, \bibinfo
  {author} {\bibfnamefont {J.}~\bibnamefont {Brown}}, \bibinfo {author}
  {\bibfnamefont {P.}~\bibnamefont {Brune}}, \bibinfo {author} {\bibfnamefont
  {K.}~\bibnamefont {Buschelman}}, \bibinfo {author} {\bibfnamefont
  {L.}~\bibnamefont {Dalcin}}, \bibinfo {author} {\bibfnamefont
  {A.}~\bibnamefont {Dener}}, \bibinfo {author} {\bibfnamefont
  {V.}~\bibnamefont {Eijkhout}}, \bibinfo {author} {\bibfnamefont {W.~D.}\
  \bibnamefont {Gropp}}, \bibinfo {author} {\bibfnamefont {D.}~\bibnamefont
  {Karpeyev}}, \bibinfo {author} {\bibfnamefont {D.}~\bibnamefont {Kaushik}},
  \bibinfo {author} {\bibfnamefont {M.~G.}\ \bibnamefont {Knepley}}, \bibinfo
  {author} {\bibfnamefont {D.~A.}\ \bibnamefont {May}}, \bibinfo {author}
  {\bibfnamefont {L.~C.}\ \bibnamefont {McInnes}}, \bibinfo {author}
  {\bibfnamefont {R.~T.}\ \bibnamefont {Mills}}, \bibinfo {author}
  {\bibfnamefont {T.}~\bibnamefont {Munson}}, \bibinfo {author} {\bibfnamefont
  {K.}~\bibnamefont {Rupp}}, \bibinfo {author} {\bibfnamefont {P.}~\bibnamefont
  {Sanan}}, \bibinfo {author} {\bibfnamefont {B.~F.}\ \bibnamefont {Smith}},
  \bibinfo {author} {\bibfnamefont {S.}~\bibnamefont {Zampini}}, \bibinfo
  {author} {\bibfnamefont {H.}~\bibnamefont {Zhang}}, \ and\ \bibinfo {author}
  {\bibfnamefont {H.}~\bibnamefont {Zhang}},\ }\href
  {https://www.mcs.anl.gov/petsc} {\emph {\bibinfo {title} {{PETS}c Users
  Manual}}},\ \bibinfo {type} {Tech. Rep.}\ \bibinfo {number} {ANL-95/11 -
  Revision 3.12}\ (\bibinfo  {institution} {Argonne National Laboratory},\
  \bibinfo {year} {2019})\BibitemShut {NoStop}%
\bibitem [{\citenamefont {Balay}\ \emph {et~al.}(1997)\citenamefont {Balay},
  \citenamefont {Gropp}, \citenamefont {McInnes},\ and\ \citenamefont
  {Smith}}]{petsc-efficient}%
  \BibitemOpen
  \bibfield  {author} {\bibinfo {author} {\bibfnamefont {S.}~\bibnamefont
  {Balay}}, \bibinfo {author} {\bibfnamefont {W.~D.}\ \bibnamefont {Gropp}},
  \bibinfo {author} {\bibfnamefont {L.~C.}\ \bibnamefont {McInnes}}, \ and\
  \bibinfo {author} {\bibfnamefont {B.~F.}\ \bibnamefont {Smith}},\ }in\
  \href@noop {} {\emph {\bibinfo {booktitle} {Modern Software Tools in
  Scientific Computing}}},\ \bibinfo {editor} {edited by\ \bibinfo {editor}
  {\bibfnamefont {E.}~\bibnamefont {Arge}}, \bibinfo {editor} {\bibfnamefont
  {A.~M.}\ \bibnamefont {Bruaset}}, \ and\ \bibinfo {editor} {\bibfnamefont
  {H.~P.}\ \bibnamefont {Langtangen}}}\ (\bibinfo  {publisher}
  {Birkh{\"{a}}user Press},\ \bibinfo {year} {1997})\ pp.\ \bibinfo {pages}
  {163--202}\BibitemShut {NoStop}%
\bibitem [{\citenamefont {Baraldo}\ \emph {et~al.}(1999)\citenamefont
  {Baraldo}, \citenamefont {Hosoya},\ and\ \citenamefont
  {Nakamura}}]{Baraldo:1999ny}%
  \BibitemOpen
  \bibfield  {author} {\bibinfo {author} {\bibfnamefont {C.}~\bibnamefont
  {Baraldo}}, \bibinfo {author} {\bibfnamefont {A.}~\bibnamefont {Hosoya}}, \
  and\ \bibinfo {author} {\bibfnamefont {T.~T.}\ \bibnamefont {Nakamura}},\
  }\href {\doibase 10.1103/PhysRevD.59.083001} {\bibfield  {journal} {\bibinfo
  {journal} {Phys. Rev.}\ }\textbf {\bibinfo {volume} {D59}},\ \bibinfo {pages}
  {083001} (\bibinfo {year} {1999})}\BibitemShut {NoStop}%
\bibitem [{\citenamefont {Cusin}\ and\ \citenamefont
  {Lagos}(2019)}]{Cusin:2019rmt}%
  \BibitemOpen
  \bibfield  {author} {\bibinfo {author} {\bibfnamefont {G.}~\bibnamefont
  {Cusin}}\ and\ \bibinfo {author} {\bibfnamefont {M.}~\bibnamefont {Lagos}},\
  }\href@noop {} {\  (\bibinfo {year} {2019})},\ \Eprint
  {http://arxiv.org/abs/1910.13326} {arXiv:1910.13326 [gr-qc]} \BibitemShut
  {NoStop}%
\bibitem [{\citenamefont {Chesler}\ and\ \citenamefont
  {Loeb}(2017)}]{Chesler:2017khz}%
  \BibitemOpen
  \bibfield  {author} {\bibinfo {author} {\bibfnamefont {P.~M.}\ \bibnamefont
  {Chesler}}\ and\ \bibinfo {author} {\bibfnamefont {A.}~\bibnamefont {Loeb}},\
  }\href {\doibase 10.1103/PhysRevLett.119.031102} {\bibfield  {journal}
  {\bibinfo  {journal} {Phys. Rev. Lett.}\ }\textbf {\bibinfo {volume} {119}},\
  \bibinfo {pages} {031102} (\bibinfo {year} {2017})},\ \Eprint
  {http://arxiv.org/abs/1704.05116} {arXiv:1704.05116 [astro-ph.HE]}
  \BibitemShut {NoStop}%
\bibitem [{\citenamefont {Belgacem}\ \emph {et~al.}(2019)\citenamefont
  {Belgacem} \emph {et~al.}}]{Belgacem:2019pkk}%
  \BibitemOpen
  \bibfield  {author} {\bibinfo {author} {\bibfnamefont {E.}~\bibnamefont
  {Belgacem}} \emph {et~al.} (\bibinfo {collaboration} {LISA Cosmology Working
  Group}),\ }\href {\doibase 10.1088/1475-7516/2019/07/024} {\bibfield
  {journal} {\bibinfo  {journal} {JCAP}\ }\textbf {\bibinfo {volume} {1907}},\
  \bibinfo {pages} {024} (\bibinfo {year} {2019})},\ \Eprint
  {http://arxiv.org/abs/1906.01593} {arXiv:1906.01593 [astro-ph.CO]}
  \BibitemShut {NoStop}%
\end{thebibliography}%

\end{document}